\documentclass[12pt]{article}

\usepackage{url}
\usepackage{amsmath,amssymb,amsthm} 
\usepackage{appendix} 
\usepackage{bm} 
\usepackage{mathrsfs} 
\usepackage{multirow} 
\usepackage{float} 
\usepackage{bbm} 
\usepackage{slashed} 

\usepackage{latexsym}
\usepackage{amsmath}
\usepackage{amsfonts}
\usepackage{amssymb}
\usepackage{amscd}
\usepackage{amsthm}

\usepackage{bbm}
\usepackage{fancybox}
\usepackage{cite}
\usepackage{amsmath,amsfonts,amsbsy}
\usepackage{pstricks,pst-node}
\usepackage[small,bf,hang]{caption2}
\usepackage{slashed}
\usepackage{graphicx}
\usepackage{epsfig}
\usepackage{psfrag}
\usepackage{comment}
\usepackage[normalem]{ulem}
\usepackage{tikz}
\newtheorem{theorem}{Theorem}

\newtheorem{definition}{Definition}
\newtheorem{proposition}{Proposition}

\usepackage{float}
\usepackage{enumerate}
\psset{unit=1.3cm,linewidth=.5pt,radius=.2} 

\usepackage{multirow}           
\usepackage{float}             
\usepackage{lscape}             
\usepackage{bm}

\newcommand{\beq}{\begin{equation}}
\newcommand{\eeq}{\end{equation}}
\newcommand{\bi}{\begin{itemize}}
\newcommand{\ei}{\end{itemize}}
\newcommand{\bt}{\begin{tabular}}
\newcommand{\et}{\end{tabular}}
\newcommand{\bc}{\begin{center}}
\newcommand{\ec}{\end{center}}

\newcommand{\ket}[1]{|#1\rangle}

\newcommand{\be}{\begin{equation}}
\newcommand{\ee}{\end{equation}}
\newcommand{\baa}{\begin{equation}\begin{aligned}}
\newcommand{\eaa}{\end{aligned}\end{equation}}
\newcommand{\bea}{\begin{eqnarray}}
\newcommand{\eea}{\end{eqnarray}}
\newcommand{\ba}{\begin{array}}
\newcommand{\ea}{\end{array}}

\def\bbox{{\,\lower0.9pt\vbox{\hrule \hbox{\vrule height 0.2 cm
\hskip 0.2 cm \vrule height 0.2 cm}\hrule}\,}}
\newcommand{\dsl}{\pa \kern-0.5em /}

\newcommand{\bC}{\mathbb C}
\newcommand{\bD}{\mathbb D}

\newcommand{\bH}{\mathbb H}

\newcommand{\bJ}{\mathbb J}

\newcommand{\bM}{\mathbb M}
\newcommand{\bN}{\mathbb N}

\newcommand{\bR}{\mathbb R}

\newcommand{\bZ}{\mathbb Z}

\newcommand{\ospr}{\mathfrak{osp}(4^*|4)}

\def\tr{{\rm tr}}

\makeatletter \@addtoreset{equation}{section} \makeatother

\title{A Superconformal Index for HyperK\"{a}hler Cones}

\author{Alec E. Barns-Graham and Nick Dorey\\
Department of Applied Mathematics and Theoretical Physics, \\
University of Cambridge, \\
Cambridge, CB3 0WA, UK \\
{\tt  n.dorey, a.e.barns-graham@damtp.cam.ac.uk}\\}

\begin{document}
\maketitle
\abstract{We define an index for $\mathfrak{osp}(4^{*}|4)$
superconformal quantum mechanics on a hyperK\"{a}hler cone. The index
is defined on an equivariant symplectic resolution of the cone, which
acts as a regulator. We present evidence that the index does not
depend on the choice of resolution parameters and encodes information
about the spectrum of (semi-) short representations of the
superconformal algebra of the unresolved space. In particular, there
are two types of multiplet which can be counted exactly using the
index. These correspond to holomorphic functions on the cone and to the
generators of the Borel-Moore homology on the resolved space respectively. We calculate the
resulting index by localisation for a large class of examples.}

\maketitle

\tableofcontents

\section{Introduction}

Conformal quantum mechanics refers to a
quantum mechanical system in which the Hamiltonian
is accompanied by generators for dilatations and special conformal
transformations which together form an $SO(2,1)$ symmetry group,
\cite{de:1976conformal}.
Such models are interesting in the context of holographic
duality where they may provide a boundary description of geometries
containing an AdS$_{2}$ factor such as the near horizon region of 
a black hole. Conformal quantum mechanics also arises as  
the discrete light-cone quantisation (DLCQ) of higher-dimensional CFTs, \cite{Aharony:1997matrix,Aharony:1997light}.

\paragraph{}
The maximally supersymmetric case we study here,
arises for a quantum mechanical
$\sigma$-model with a hyperK\"{a}hler target manifold. As usual
the resulting theory has $\mathcal{N}=(4,4)$ supersymmetry, \cite{Alvarez:1981geometrical}. The target
space also satisfies the
criterion for $\mathfrak{so}(2,1)$ conformal invariance if it admits a
closed homothety in the sense of \cite{Michelson:2000geometry}. Such manifolds
are known as {\em
  hyperK\"{a}hler cones}.
Provided the homothety is
also triholomorphic\footnote{For brevity, this condition is implied
  whenever we use the term
    `hyperK\"{a}hler cone' in the
  following.}, the theory acquires a larger $\mathfrak{osp}(4^{*}|4)$
superconformal symmetry, whose bosonic subalgebra includes $\mathfrak{so}(2,1)$
conformal transformations together with an $\mathfrak{su}(2)\oplus
\mathfrak{usp}(4)$ algebra of R-symmetry transformations, \cite{Singleton:2014}.
Large classes of  hyperK\"{a}hler cones obeying this condition
arise as the Higgs branches of supersymmetric gauge theories in higher
dimensions. This provides a physical context for the
corresponding $\sigma$-models; compactifying the spatial
dimensions of the gauge theory on a torus, the resulting theory flows
to superconformal quantum mechanics in the IR.
In mathematical terms the Higgs branch of a
supersymmetric gauge theory with eight supercharges
corresponds to a particular hyperK\"{a}hler quotient of
flat space. For quiver gauge theories, the resulting spaces also have
an algebro-geometric description and are known
as Nakajima quiver varieties. As we discuss below, these are singular
spaces and a resolution of the singularities is needed in order to
properly define supersymmetric quantum mechanics. 
\paragraph{}
One particularly interesting model in this class is
the ADHM quiver corresponding to the moduli space of
Yang-Mills instantons on $\mathbb{R}^{4}$. This model is believed to
provide a DLCQ description of the $(2,0)$ theory
in six dimensions. This identification is consistent with the fact
that $\mathfrak{osp}(4^{*}|4)$ is a factor in the subalgebra of the $(2,0)$
superconformal algebra in six dimensions which is left unbroken by
compactification of a null direction. It further suggests the existence of a
an $\mathfrak{osp}(4^{*}|4)$ invariant quantum mechanics on the
instanton moduli space with a spectrum consisting of lowest-weight unitary
representations of this algebra. These states should be the ones
arising from the branching of unitary representations of the full
$(2,0)$ superconformal algebra onto those of the
lightcone subalgebra. In \cite{Aharony:1997light}, Aharony et al 
regulated the ADHM $\sigma$-model by suitably
resolving the singularities of the
instanton moduli space. Any such deformation 
inevitably breaks superconformal invariance. However, the authors of
\cite{Aharony:1997light} argued that the resolution corresponds to a UV regulator for
six-dimensional spacetime theory and also suggested a limiting procedure to
extract the observables of conformal quantum mechanics from those
of the regulated theory. 
\paragraph{}
In this paper we will proceed in a similar
way for more general models in the same class and test the
hypothesis that 
{\em every} hyperK\"a{h}ler cone defines an  $\mathfrak{osp}(4^{*}|4)$
invariant superconformal quantum mechanics with a spectrum consisting
of lowest-weight unitary representations. For a large class of such
models there is a natural choice of regulator; 
we replace the hyperK\"{a}hler cone by its {\em
equivariant symplectic resolution}\footnote{The resolution is in fact a projective symplectic
resolution, which necessarily has the property of being equivariant.}. For cones which arise as Higgs
branches of quiver gauge theories, such resolutions are
provided by turning on 
generic values of the real Fayet-Illiopoulos parameters in the gauge
theory. In the case of the ADHM quiver this coincides with the
resolution considered in \cite{Aharony:1997light}. 
\paragraph{}
As for superconformal algebras in higher dimensions, representations of
$\mathfrak{osp}(4^{*}|4)$ are naturally classified as (semi-) short or
long depending on whether the dimensions of the primary state saturate
a BPS bound. In recent work, \cite{Singletonthesis} (see also \cite{Dorey:2018})  Singleton
defined a superconformal index which counts the (semi-) short multiplets
modulo the possiblity of recombination into long ones. As well as 
 encoding the dimensions and R-charges of superconformal
primaries, the index also grades states
according to their quantum numbers under the global symmetry group of
the theory. In the case of flat $\mathbb{C}^{2n}$, the states
contributing to the index are in one-to-one correspondence with
polynomial-valued holomorphic forms on the target space. 
\paragraph{}
In the general case we replace the singular cone by its equivariant symplectic 
resolution. Although the full superconformal 
algebra is broken by the resolution, the Cartan generators of the
little-group associated with the superconformal index all remain
unbroken as do the triholomorphic isometries of the hyperK\"{a}hler
cone. In the following, we will argue that the superconformal index can be
identified with an appropriate index of the resolved space, namely a certain
equivariant Euler character of sheaf cohomology. We will show
that the index defined in this way correctly captures the
expected features of superconformal quantum mechanics on the
underlying singular space and present evidence that it does not depend
on the choice of symplectic resolution. 
\paragraph{}
To define the index we need to pass to the algebraic description of
the hyperK\"{a}hler cone $\mathcal{X}$ and its equivariant
symplectic resolution
$\tilde{\mathcal{X}}$ as a complex variety. 
In this context we
consider the sheaf $A^p(\tilde{\mathcal{X}})$ of $p$-forms on 
$\tilde{\mathcal{X}}$ and the associated sheaf cohomology $H^{*}$ 
in the Zariski topology. In section \ref{Superconformal index section}, we argue that the superconformal index can be identified with an equivariant analog of the Hirzebruch genus in this complex. The resulting index $\mathcal{Z}(\mathcal{X})$
for a hyperK\"{a}hler cone $\mathcal{X}$ of
quaternionic dimension $d_{H}$,
is a function of two
fugacities $\tau$ and $y$ which can be identified with the Cartan generators of
the $SU(1|2)$ little group preserved by the BPS shortening condition
of the superconformal algebra as well as additional fugacities
$Z=\{z_{i}\}$, corresponding to global symmetries.
\baa\label{Definition of superconformal index1}
\mathcal Z({\mathcal
X};\tau,y,Z):&=\sum_{p,q=0}^{2d_H}(-)^{p+q}\left(\frac
y\tau\right)^{p-d_H}\tr_{H^q(\tilde{\mathcal X},A^p(\tilde{\mathcal
X}))}\left(\tau^{R}\prod_iz_i^{\mathcal J_i}\right)\,.
\eaa
Geometrically, the powers of
$y/\tau$ appearing in the index correspond to
holomorphic degree, $p$, in sheaf cohomology while those of $\tau$
record the grade $R$ under the $\mathbb{C}^{\times}$ action preserved by
the symplectic resolution. The global symmetry fugacities $z_{i}$, grade the
cohomology classes according to their charge under
the triholomorphic isometries of
$\tilde{\mathcal{X}}$.
\paragraph{}
For all the cases we
study in this paper, the index can be localised at fixed points of a
group action on the manifold and calculated exactly. Applying the
relevant localisation theorems we find,
\baa\label{SuperconformalIndexDerivation1}
\mathcal Z(\mathcal X)
&=\left(\frac\tau y\right)^{d_H}\sum_{x\in\tilde{\mathcal
X}^T}\text{PE}\left[\left(1-\frac
y\tau\right)\text{ch}_T(T^*_x\tilde{\mathcal X};\tau,Z)\right]\,.
\eaa
Here the sum is over fixed points $x$ of $\tilde{\mathcal{X}}$
under the torus $T$ generated by the isometries $\mathcal J_{i}$ and
$\mathbb{C}^{\times}$. In the above formula $T^*_x\tilde{\mathcal X}$
is the tangent space to the manifold at the fixed point $x$
considered as a $T$-module and $\text{ch}_{T}$ denotes the
corresponding character. Finally, $\text{PE}$ denotes the plethystic
exponential. For the full definitions see Section 2.1.
\paragraph{}
By construction, the
index has no continuous dependence on the resolution parameters. 
However, it is known that hyperK\"{a}hler cones can have inequivalent
symplectic resolutions corresponding to different chambers in the
parameter space separated by walls of codimension one, \cite{Nakajima:1994}. One might
therefore worry that the index we define depends on the choice of
chamber. In the following, we will present evidence that the index is
in fact independent of this choice. The evidence consists of several
different limits and specialisations of the index for which this can
be proven. Finally, we perform some explicit calculations for a generic
quiver of low dimension.       
\paragraph{}
As in higher dimensions, the expansion
of the index in characters of the little group provides information about the
spectrum of $\mathfrak{osp}(4^{*}|4)$ multiplets. 
Although the index
typically provides only a lower bound on the spectrum, there are two types
of multiplets which we can count precisely. Both are related to the
geometry of the target space. The first are the 
$1$/$2$-BPS short multiplets which we show are in one-to-one
correspondence with the generators of Borel-Moore homology of the resolved
space. These states cannot be lifted and the index counts them without
sign. The second are a particular type of semi-short multiplet of 
$\mathfrak{osp}(4^{*}|4)$ which
are in one-to-one correspondence with holomorphic functions on the
target space. The partition function for these two classes of
protected states are related to the Hilbert series and Poincare
polynomial of $\tilde{\mathcal{X}}$ respectively. In fact, both these
partition functions arise as particular limits of the superconformal
index. As we explain in Section 3, the Hilbert series arises
in the $y\rightarrow 0$ limit of $\mathcal{Z}(\tilde{\mathcal{X}})$
while the Poincare polynomial in Borel-Moore homology appears in the
$\tau\rightarrow\infty$ limit with $y/\tau$ held fixed. The full index
interpolates between these two quantities and encodes a lower bound on
the degeneracies of the other (semi-)short multiplets of 
$\mathfrak{osp}(4^{*}|4)$ in each irreducible 
representation of the global symmetry. 
\paragraph{}
In the special case of
the ADHM quiver, the index (\ref{Definition of superconformal index1})
coincides with the corresponding
instanton contribution to the Nekrasov partition function of a certain
five-dimensional supersymmetric gauge theory, and agrees with an
earlier proposal for a superconformal index for ADHM quantum
mechanics, \cite{Kim:11}. In this case, we confirm 
the identification of $1$/$2$-BPS multiplets 
with compactly supported cohomology classes which was anticipated in
\cite{Aharony:1997light,Kim:11}. This identification reproduces the known
spectrum of chiral primaries of the $(2,0)$ theory corresponding to
the Kaluza-Klein modes of eleven dimensional supergravity on
$AdS_{7}\times S^{4}$. The semi-short multiplets of
$\mathfrak{osp}(4^{*}|4)$ should be related to the spectrum of $1$/$8$-BPS states of the
$(2,0)$ theory. The resulting bound from the index
should be relevant for counting the
microstates of supersymmetric black holes in $AdS_{7}$.
\paragraph{}
The superconformal index defined in this paper
has a number of other interesting properties.
HyperK\"{a}hler cones typically contain many cones of lower dimension,
which appear as fixed points of symmetries.
The superconformal index is stable under this reduction in the
following sense; the index of the fixed point
submanifold is obtained simply by taking the fugacity of the
corresponding symmetry to zero in the index of the original space.
This connection provides an efficient way of generating the
superconformal indices of all quivers varieties of type $A_{n}$ and
$\hat{A}_{n}$, as these all appear as fixed points in the instanton moduli
space \cite{Nakajima:11}. The connection to the Poincare polynomial,
allows us to write an explicit formula for the partition function of $1$/$2$-BPS states for superconformal quantum
mechanics on any Nakajima quiver variety. There are interesting
connections with the work of Nakajima, who constructed an action of a
quantum affine algebra on the Borel-Moore homology of these spaces \cite{Nakajima:1994,Nakajima:1998}.
Indeed the resulting generating function can be
identified with graded character of a specific module
of this algebra.
\paragraph{}
The paper is organised as follows. In Section 2, we will review
$\mathfrak{osp}(4^{*}|4)$ superconformal quantum mechanics on a
hyperK\"ahler cone. We also review the
symplectic resolution of the cone and define a regulated
superconformal index. Finally, we give a general formula for the index
via localisation to the fixed points of a group action.
In Section 3, we discuss various properties of the index including the
limiting behaviour mentioned above. To show that our interpretation of
$\mathcal{Z}$ as a superconformal index is consistent we compare our
formula (\ref{SuperconformalIndexDerivation1}) with the index
evaluated on a general spectrum of unitary lowest weight
representations of $\mathfrak{osp}(4^{*}|4)$. This comparison provides several
non-trivial tests of our proposal. Finally, in Section 4, we apply
the fixed point formula to compute the index in numerous cases.

\section{The superconformal index}\label{Superconformal index section}

Following \cite{Singleton:2014,Singletonthesis}, we will
study supersymmetric quantum mechanics on a hyperK\"ahler
manifold. As usual, the Hamiltonian $\mathbb{H}$ of the system is
identified with the Laplacian acting on forms and the hyperK\"{a}hler
condition gives an $\mathcal{N}=(4,4)$ supersymmetry algebra.
We specialize further to the case of a hyperK\"{a}hler cone (see
\cite{de2001:hypermultiplets}),
\begin{definition}
For $\mathcal X$ a hyperK\"ahler manifold, it is a \emph{hyperK\"ahler cone} if
it has a homothetic conformal Killing vector field $V_D$,
\be
\mathcal L_{V_D}g=2g\,.
\ee
\end{definition}
We define $d_H:=\frac14\dim_\bR\mathcal X\in\bZ_{\geqslant0}$. The
homothety $D$ yields, a dilatation operator $\mathbb{D}$ and also
implies the existence of a scalar function $K$. The corresponding
operator $\mathbb{K}$ together with $\mathbb{H}$ and $\mathbb{D}$
generates an $\mathfrak{so}(2,1)$ conformal algebra. As the space
$\mathcal{X}$ is necessarily non-compact, these operators act on the
Hilbert space of $L^{2}$ normalisable forms. In the simplest case of
flat space one finds a discrete spectrum\footnote{In conventional quantum
mechanics on $X$, one would instead diagonalise $\mathbb{H}$ in the slightly
larger Hilbert space of plane-wave normalisable forms leading to a
continuous spectrum of scattering states.} for the dilatation operator
$\mathbb{D}$
corresponding to a set of unitary representations of
$\mathfrak{so}(2,1)$.  More generally, a standard argument shows that
$\mathbb{D}$ is isospectral to $\mathbb{L}_{0}=\mu^{-1}\mathbb{H}+\mu
\mathbb{K}$ for arbitrary $\mu$. The addition of $\mathbb{K}$ to the
Hamiltonian provides
a harmonic potential on $\mathcal{X}$ which should lead to a discrete
spectrum for $\mathbb{L}_{0}$ and thus for $\mathbb{D}$.
 \paragraph{}
If the homothety $D$ is triholomophic then the $\mathfrak{so}(2,1)$
conformal algebra combines with $\mathcal{N}=(4,4)$ supersymmetry
give a larger $\mathfrak{osp}(4^*|4)$ superconformal algebra
\cite{Singleton:2014}. In partcular, in this reference, Singleton gave
an explict presentation of the
generators of $\mathfrak{osp}(4^*|4)$ acting on the the space of forms on
$\mathcal{X}$. If $\mathcal{X}$ is smooth, this algebra is therefore
realised in the spectrum of supersymmetric quantum mechanics on the
manifold. In the case of a flat target space $\mathbb{R}^{4n}$,
the model can be solved exactly and one finds a
spectrum consisting of postive energy, unitary
irreducible representations of the superconformal algebra.
Unfortunately, this is
the only non-singular hyperK\"{a}hler manifold which admits a
triholomorphic homethety. In particular, the
hyperK\"{a}hler cones we consider here are singular spaces and
additional input is required to define a sensible model.
\paragraph{}
In this
paper, we will consider the hypothesis that there is an $\mathfrak{osp}(4^*|4)$
invariant theory associated with
each hyperK\"{a}hler cone $\mathcal{X}$ which in particular has
a spectrum consisting of a countable set of positive-energy,
irreducible unitary representations of this algebra. We will now
review the classification of these representations and the
corresponding superconformal index.

\paragraph{}
The bosonic subalgebra of the superconformal algebra is
\be
\mathfrak
g_B=\mathfrak{so}(2,1)\oplus\mathfrak{su}(2)\oplus\mathfrak{usp}(4)\,.
\ee
The Cartan subalgebra of $\mathfrak g_B$ is generated by $\bD,\bJ_3,\bM$ and
$\bN$,
with $\bJ_3$ the Cartan generator of $\mathfrak{su}(2)$, and $\bM$ and $\bN$
the
Cartan generators of $\mathfrak{usp}(4)$. The weight lattice is generated in
the orthogonal basis as
$\epsilon_1\bZ\oplus\epsilon_2\bZ\oplus\delta_1\bZ\oplus\delta_2\bZ$, defined
such that if $v_\lambda$ has eigenvalues $(\Delta,-2j,-m,-n)$ under
$(\bD,2\bJ_3,\bM,\bN)$,  then $v_\lambda$ has weight
\be
\lambda=\frac\Delta2(\epsilon_1+\epsilon_2)-j(\epsilon_1-\epsilon_2)-m\delta_1-n\delta_2\,.
\ee
 In order for $v_\lambda$ to be a lowest weight vector for a unitary
irreducible representation of $\mathfrak{osp}(4^*|4)$, it is necessary that
$\Delta\geqslant0$, $(2j,m,n)\in\bZ_{\geqslant0}^3$ and $m\geqslant n$. On a generic state, $|\psi\rangle$, corresponding to a $(p,q)$-form on $\mathcal X$, we have
\be
\bM|\psi\rangle=(q-d_H)|\psi\rangle\,,\quad\bN|\psi\rangle=(p-d_H)|\psi\rangle\,.
\ee

\paragraph{}

As for superconformal algebras in higher dimension,
the unitary irreducible lowest weight
representations are classified by the value of the lowest weight. The
lowest weight state is annhilated by the lowering operators of
$\mathfrak{osp}(4^{*}|4)$. In
addition there are BPS bounds on the dimension
$\Delta$.  If $\Delta$ saturates the
bound, then the lowest weight vector is annihilated by one or more
additional generator. In the work \cite{Singletonthesis}, Singleton found a
full
classification of unitary irreducible lowest weight representations,

\begin{theorem}  Unitary, irreducible, lowest weight
representations of $\ospr$ are obtained from the Verma module generated
by the action of $\ospr$ on $\ket{\Delta,j,m,n}$, by quotienting out null
states. They come in the following types:
\begin{itemize} \item Generic `long' representations $L(\Delta,j,m,n)$
with $\Delta>2(j+m+1)$.
\item `Semishort' representations $SS(j,m,n)$ with $\Delta
=2(j+m+1)$.
\item `Short' representations $S(m,n)$ with $\Delta=2m$ and $j=0$. These
split into 1/2-BPS representations with $m=n$ and 1/4-BPS otherwise.
\end{itemize}
\end{theorem}

In \cite{Singletonthesis}, Singleton, defined
a superconformal index which receives
contributions solely from the short and semi-short
representations. The index can be used to count the (semi-)short
representations up to the possibility of recombination into long
multiplets. To define the index we pick a supercharge $q$ and its
conjugate, $s=q^{\dagger}$ such that,
\be
\{q,s\} \,=\, \mathcal{H} =\frac12\mathbb{L}_{0}+\mathbb{J}_{3}+\mathbb{M}\,.
\ee
Thus $\mathcal{H}$ has eigenvalues,
\be
E:=\frac12\Delta-j-m\,.
\ee
Each (semi)-short multiplet contains states which are
annihilated by $\mathcal{H}$. Assuming a discrete spectrum,
these states are in one-to-one correspondence with the
cohomology classes of the supercharge $s$.
The choice of BPS bound
breaks the full superconformal algebra down to the
$\mathfrak{su}(1|2)$ subalgebra commuting with $q$ and $s$. This subalgebra has
Cartan generators $\mathbb{T}=-(\mathbb{M}+ 2\mathbb{J}_{3})$ and $\mathbb{N}$.
The superconformal index counts states saturating the bound, graded by
their charges under $\mathbb{T}$ and $\mathbb{N}$ and  any
additional mutually commuting global symmetry generators $\{\mathcal J_{i}\}$. The fermion number is given by $F=\bM+\bN$.
The resulting index is given as,
\be
\mathcal{I}\left(t,y,Z\right)={\rm Tr}\left[ (-1)^{F}e^{-\beta
\mathcal{H}}\, \tau^{\mathbb{T}}y^{\mathbb{N}}\prod_iz_i^{\mathcal
J_{i}}\right]\,.
\ee
Assuming a discrete spectrum, the superconformal index is invariant
under all deformations of the system which preserve the supercharges
$q$ and $s$. In particular, the index is independent of the parameter
$\beta$.
\paragraph{}
The states with $E=0$ in each
(semi-)short multiplet of $\mathfrak{osp}(4^*|4)$ transform in representations
of the ``little group'' $SU(1|2)$ their contribution to the index is the
corresponding character. The $1/2$- and $1/4$-BPS  short representations
$S(m,m)$ and
$S(m,n)$ with $m>n\geq 0$ have characters,
\baa\label{SU21characters}
\mathcal{I}_{m,m}\left(\tau,y\right) & = \tau^{m}\left[\chi_{m}(y)-\tau
  \chi_{m-1}(y)\right] \,, \\
{\mathcal I}_{m,n}\left(\tau,y\right) & = \tau^{m}\left[(1+\tau^{2})
\chi_{n}(y)-\tau\left(\chi_{n+1}(y)+\chi_{n-1}(y)\right)\right]\,,
\eaa
where $\chi_{n}(y)$ is the character of the spin $n/2$ representation
of $\mathfrak{su}(2)$;
\be
\chi_{n}(y) =  y^{n}+y^{n-2}+\ldots +y^{-n} \,.
\ee
$\chi_{0}(y)=1$ and we adopt the convention that
$\chi_{-1}(y)=0$. Similarly, the semi-short representation $SS(j,m,n)$
yields the character,
\baa
\mathcal{I}_{j,m,n}\left(\tau,y\right)&=\tau^{m+2j+2}\left[(1+\tau^{2})
\chi_{n}(y)-\tau\left(\chi_{n+1}(y)+\chi_{n-1}(y)\right)\right]\\
&=\tau^{2j+2}
{\mathcal I}_{m,n}\left(\tau,y\right) \,.
\eaa

\paragraph{}
For a generic $\mathfrak{osp}(4^{*}|4)$ invariant theory, with
(semi-) short spectrum,
\baa\label{Spectrum}
\mathcal{S} =&\left(\bigoplus_{m= 0}^{d_H}\,\,
N^{(m,m)}S(m,m)\right)\,\oplus\left( \bigoplus_{m=n+1 }^{d_H}\bigoplus_{n=0}^{d_H-1}\,\,
N^{(m,n)}S(m,n)\right) \\
&\oplus\, \left(\bigoplus_{m= n}^{d_H-1}\bigoplus_{n=0}^{d_H-1}\bigoplus_{ j\in\frac12\bZ_{\geqslant0}}
\,\, N^{(j,m,n)}SS(j,m,n)\right) \,.
 \eaa
The upper bounds in the direct sums in equation (\ref{Spectrum}) come  from the geometric constraint that the holomorphic and antiholomorphic degrees $p$ and $q$ of any form are both bounded by $2d_H$. 
\paragraph{} 
 
The superconformal index can be expressed in
terms\footnote{The polynomials $N^{(m,n)}$ and $N^{(j,m,n)}$ are characters of the global symmetry. When the global symmetry is non-Abelian, the polynomials will be invariant under the corresponding Weyl group. We will see that the $N^{(m,m)}$ have no $Z$
dependence for all cases we investigate.} of
$N^{(m,m)},N^{(m,n)},N^{(j,m,n)}\in\bZ_{\geqslant0}[Z,Z^{-1}]$ as,
\baa
\mathcal Z(\tau,y;Z) =& \sum_{m=0}^{d_H}  N^{(m,m)}
\mathcal{I}_{m,m}\left(\tau,y\right)-\sum_{m=1}^{d_H}  N^{(m,m-1)}(Z)
\mathcal{I}_{m,m-1}\left(\tau,y\right)\,\\
&+\sum_{m=n+2}^\infty \sum_{n=0}^{d_H-1} (-1)^{m-n}\,
{\tilde N}^{(m,n)}(Z)\mathcal{I}_{m,n}\left(\tau,y\right) \,,
\label{general}
\eaa
where for $m\geqslant n+2$, we have,
\be
{\tilde N}^{m,n}(Z) = N^{(m,n)}(Z)+\sum_{k=\max\{0,m-1-d_H\}}^{m-n-2} (-1)^{k+1}
N^{(k/2,m-2-k,n)}(Z)\,.
\label{ntilde}
\ee

 Given the value of the index ${\mathcal I}$ as a function of $\tau$ and $y$ it
is possible to read off the numbers $N^{(m,m)}$, $N^{(m+1,m)}$ and
$\tilde{N}^{(m,n)}$. The alternating signs in the expression for ${\tilde
N}^{m,n}$ correspond to potential cancellations between different (semi-)short
multiplets contributing to the index. In some special cases these cancellations are absent, and we can therefore uniquely determine the degeneracies of the corresponding multiplets. Specifically, we can  uniquely
determine the numbers of
1/2-BPS short multiplets $S(m,m)$, 1/4-BPS short multiplets $S(n+1,n)$, and also the semishort
representations $SS(j,d_H-1,d_H-1)$. However,
 for the other short and
semi-short multiplets of $\mathfrak{osp}(4^*|4)$ the index instead provides a
lower bound for the degeneracies of these states.

\paragraph{}
As mentioned above, Singleton constructed a geometric action of
$\mathfrak{osp}(4^*|4)$ on the space of differential forms on a
hyperK\"{a}hler cone $\mathcal{X}$.
After the change of basis from $\mathbb{D}$ to
$\mathbb{L}_{0}$, the inner product on the space of forms becomes,
\be\label{HilbertSpacedotproduct}
(\alpha,\beta)=\int_{\mathcal X}\text
d^{4d_H}x\sqrt{g}\alpha\wedge\overline\beta
e^{-\mu K}\,,
\ee
where $4d_H=\dim_\bR\mathcal X$, and $K$ is the function on
$\mathcal{X}$ corresponding to the special conformal generator
$\mathbb{K}$. In the case of flat affine space
$\mathcal{X}=\mathbb{R}^{4n}=\mathbb{C}^{2n}$
the index can be calculated easily. In this case, the representatives in each
superconformal multiplet which contribute to the index are in one to
one correspondence with the holomorphic forms on the target
space. The following analysis of the index is only
rigorously correct for the flat space case, but will suggest a
regularised definition of the index in the general case. The resulting
index is a trace over the space of forms with vanishing
$\mathcal{H}$-eigenvalue on
$\mathcal X$ graded by the triholomorphic isometries of the manifold, $G_H$, a
Lie group whose Cartan subalgebra is generated
by $\mathcal J_i$, and two Cartan elements of the little group
$\mathfrak{su}(1|2)$,
\be
\mathcal{Z}_{\mathcal{X}}=
\tr_{\Omega^*(\mathcal
X)}\left((-)^{\bM+\bN}e^{-\beta \mathcal
H}y^\bN\tau^{-\bM-2\bJ_3}\prod_iz_i^{\mathcal
J_i}\right)\,.
\ee
More precisely, the trace is over the space of
forms which are normalisable with respect to the inner product
(\ref{HilbertSpacedotproduct}).

\paragraph{}
If we choose a particular complex structure, the hyperK\"{a}hler
space $\mathcal{X}$ becomes a holomorphic symplectic manifold.
The R-symmetry on $\mathcal{X}$ yields  a holomorphic $\bC^\times$-action on
$\mathcal X$ under which the holomorphic symplectic form has charge 2. It is
generated by a vector field $V_{R}$. On flat space, for $a\in\bC^\times$ it
acts as $z\mapsto az$ and $\bar z\mapsto a^{-1}\bar z$. In general, the induced
action on forms is given by
\be
\mathcal L_{V_{R}}=-\bM+\bN-2\bJ_3\,.
\ee
From this, we see that we can write
\be
\mathcal Z_{\mathcal{X}}=\tr_{\Omega^*(\mathcal
X)}\left((-)^{\bM+\bN}e^{-\beta \mathcal{H}}\left(\frac
y\tau\right)^\bN\tau^{R}\prod_iz_i^{\mathcal J_i}\right)\,.
\ee
\paragraph{}
As mentioned above, the ground states
contributing to the index can be identified with cohomology classes of
the supercharge $s$.
From \cite{Singletonthesis} chapter 7, we know that if $\beta$ is any form on
$\mathcal X$ and $\alpha=\beta e^{-\mu K}$, then
\be
s\alpha=\frac1{\sqrt\mu}\overline\partial\beta\,e^{-\mu K}\,.
\ee
Hence, $s$ acts as a Dolbeault operator, up to the overall exponential factor
and $\mathcal{Z}_{\mathcal{X}}$ formally coincides with an index of the
corresponding Dolbeault complex.

\paragraph{}
The condition of  finite norm under (\ref{HilbertSpacedotproduct}) determines
the space of forms we should consider.  In the case of affine space
$\mathcal{X}=\bC^{2n}$ with complex coordinates $(q^i,\tilde q_i)$,
$i=1,\dots,n$, the inner product is given by (\ref{HilbertSpacedotproduct})
with $K=\sum_i(|q^i|^2+|\tilde q_i|^2)$. In this case one finds that the
Hilbert space is
\be
\{\text{states with } E=0\}\cong\bC[q,\tilde q,dq,d\tilde q]\,,
\ee
where the right hand side is the space of polynomials in the Grassmann even $q$
and $\tilde q$ and the Grassmann odd $dq$ and $d\tilde q$. In
otherwords, the states annihilated by $\mathcal{H}$ are precisely the
polynomial-valued holomorphic forms on $\mathcal{X}=\mathbb{C}^{2n}$
and the supercharge $s$ can be identified with the Dolbeault operator
$\bar{\partial}$ acting on these forms. Each monomial has a definite
$\mathbb{C}^\times$-grade corresponding to its degree which is
essentially the dimension of the corresponding state. Thus, it is natural
to work in the basis of homogeneous polynomials.
(See \cite{Singletonthesis} section 7.3 for further details of the
flat space case).
\paragraph{}
The flat space result reviewed above can also be described in a slightly
different way: the cohomology of the Dolbeault operator on polynomial-valued
holomorphic forms also coincides with the sheaf cohomology of
the affine variety $\mathcal{X}=\mathbb{C}^{2n}$ in the Zariski
topology. Thus, the analytic Dolbeault cohomology of $\bC^{2n}$ with finite
norm under the inner product is equal to the sheaf cohomology in the Zariski
topology, provided that we restrict our attention to forms of finite
$\mathbb{C}^\times$-grade. We will assume that this
identification also holds for a general hyperK\"{a}hler cone
and hence we think of $\mathcal X$ as a variety from now on and assume
that the space of $E=0$ states is given by Dolbeault cohomology in the
Zariski topology giving,
\baa
\mathcal Z_{\mathcal{X}}&=\sum_{p,q=0}^{2d_H}(-)^{p+q}\left(\frac
y\tau\right)^{p-d_H}\tr_{H^{p,q}(\mathcal
X)}\left(\tau^{R}\prod_iz_i^{\mathcal J_i}\right)\,.
\eaa
 With this, we can use Dolbeault's theorem to write
\baa
\mathcal Z_{\mathcal{X}}&=\sum_{p,q=0}^{2d_H}(-)^{p+q}\left(\frac
y\tau\right)^{p-d_H}\tr_{H^q(\mathcal X,A^p(\mathcal
X))}\left(\tau^{R}\prod_iz_i^{\mathcal J_i}\right)\,.
\eaa
Dolbeault's theorem (see
\cite{Griffiths:2014principles}) states that for $M$ a complex
manifold
\be
H^q(M;A^p(M) )=H^{p,q}(M)\,,
\ee
where the right hand side is the $\bar\partial$-Dolbeault cohomology, and the
left hand side is the sheaf cohomology of $A^p(M)$, the sheaf of holomorphic
$p$-forms on $M$.

\paragraph{}
The problem with this definition of the superconformal index is that, in most
examples, hyperK\"ahler cones are not smooth. They have singularities, notably
at the origin of the cone, but also along subspaces that intersect the origin.
The space of forms is not defined at the singularities. Only for $p=0$ is the
summand well-defined. In order to define our index, it is necessary that we
introduce some form of regularisation. In this work, we propose that the
Dolbeault cohomology of the projective symplectic resolution of $\mathcal X$ is
the appropriate regularisation. As above,  we specifically mean the Dolbeault
cohomology with respect to the Zariski topology, where this restriction from
the analytic topology to the Zariski topology is a consequence of the finite
norm restriction under the inner product (\ref{HilbertSpacedotproduct}).
\paragraph{}
We define a \emph{projective symplectic resolution}. For the definition of
words such as proper, projective etc. see \cite{Hartshorne:2013algebraic}. A
symplectic variety $\mathcal X$, is a variety, with an open set of smooth
points $\mathcal X^{\text{reg}}$ on which is defined a holomorphic symplectic
2-form.
\begin{definition}
For $\mathcal X$ a symplectic variety, a \emph{resolution} of $\mathcal X$ is a
proper surjective morphism, $\pi:\tilde{\mathcal X}\to \mathcal X$,  such that
$\tilde{\mathcal X}$ is smooth, and $\pi^{-1}(\mathcal
X^{\text{\emph{reg}}})\to \mathcal X^{\text{\emph{reg}}}$ is an isomorphism. If
$\pi$ is a projective morphism, then this a \emph{projective resolution}.

A \emph{symplectic resolution} is one where $\pi^*\omega$, the pullback of the
symplectic form on $\mathcal X^{\text{reg}}$, the open set of smooth points in
$\mathcal X$, can be extended to a symplectic form on all of $\tilde{\mathcal
X}$.
\end{definition}
We will assume that a projective symplectic resolution of $\mathcal X$,
$\tilde{\mathcal X}$, exists.
This is certainly the case for a large class of hyperK\"{a}hler cones
$\mathcal X$ corresponding to Nakajima quiver varieties. We
briefly recall their definition.

A quiver $\Gamma=(V,\Omega)$ is a set $V$ of vertices and $\Omega$ a set of
arrows, $h=(i,j)\in\Omega$ corresponds to $i\to j$ for $i,j\in V$, we write
in$(h)=i$, out$(h)=j$. We then provide the data $k\in\bZ_{>0}^V,$
$N\in\bZ_{\geqslant0}^V$; and
$\zeta\in\bR^{3V}$. With this, we define the affine space of complex matrices
\baa
M\equiv M(k,N)&:=\bigoplus_{(i,j)\in
\Omega}\text{Hom}\left(\bC^{k_i},\bC^{k_j}\right)\oplus\text{Hom}\left(\bC^{k_j},\bC^{k_i}\right)\\
&\phantom{:=}\oplus\bigoplus_{i\in
V}\text{Hom}\left(\bC^{N_i},\bC^{k_i}\right)\oplus
\text{Hom}\left(\bC^{k_i},\bC^{N_i}\right)\\
&\cong (\bC^2)^{\sum_{(i,j)\in \Omega}k_ik_j+\sum_{i\in V}k_iN_i}\,.
\eaa
This space is hyperK\"ahler. Elements $(X,\tilde
X,q,\tilde q)\equiv(X_{ij},\tilde
X_{ij},q_i,\tilde q_i)_{i,j}\in M$ transform under $g\in G\equiv
G_k=\prod_iGL(\bC^{k_i})$ as
\be
(X_{ij},\tilde X_{ij},q_i,\tilde q_i)\mapsto (g_jX_{ij}g_i^{-1},g_i\tilde
X_{ij}g_j^{-1},g_iq_i,\tilde q_ig_i^{-1})\,.
\ee
This action is smooth (except for the zeroes), Hamiltonian, isometric and
triholomorphic. So, we have three moment maps
\baa
\mu_\bR&:=[X,X^\dagger]+[\tilde X,\tilde X^\dagger]+qq^\dagger-\tilde q^\dagger
\tilde q\in \prod_{a\in V}\mathfrak u(k_a)\,,\\
\mu_\bC&:=[X,\tilde X]+q\tilde q\in\prod_{a\in V}\mathfrak{gl}(k_a)\,.
\eaa
We  define the \emph{Nakajima quiver variety} as
\be
\mathfrak M_{\zeta_\bR,\zeta_\bC}\equiv \mathfrak
M_{\zeta_\bR,\zeta_\bC}(k,N):=\mu_\bR^{-1}(\zeta_\bR)\cap\mu_{\bC}^{-1}
(\zeta_\bC)/G\,.
\ee
A Nakajima quiver variety is hyperK\"{a}hler
by virtue of the hyperK\"{a}hler quotient
construction.
\paragraph{}
Such varieties arise as
the Higgs branch moduli space of eight supercharge quiver
gauge theories, where $\vec\zeta$ correspond to the Fayet-Iliopoulos parameters
in the field theory Lagrangian.  Setting these parameters to zero, the singular
or unresolved Nakajima quiver   $\mathcal X=\mathfrak M_0:=\vec\mu^{-1}(0)/G$
is a hyperK\"{a}hler cone with a triholomorphic homothety and  thus gives rise
to $\mathfrak{osp(4^*|4)}$ superconformal quantum
mechanics.
\paragraph{}
Often, there are values of the
level set $\vec\zeta\in\bR^3\otimes\pi_1(G)^\vee\cong\bR^{3V}$ such that
$\mathfrak M_{\vec\zeta}:=\vec\mu^{-1}(\vec\zeta)/G$ is smooth, thus
providing a resolution of the singularity.
The values of $\zeta$ such that $\mathfrak M_{\vec\zeta}$ is smooth are known
as generic values. Either there are no generic values, or they form a subset of
$\bR^3\otimes\pi_1(G)^\vee$ whose complement is codimension 3. The
$\bC^\times$-action generated by $R$ that we grade by is only defined if
$\zeta_\bC=0$. In this case, for $(\zeta_\bR,0)$ is generic, the resulting
manifold,
$\mathfrak M_{\zeta_\bR,0}$, is a
projective symplectic resolution of the singular space $\mathfrak M_0$. From
\cite{Ginzburg:2009lectures}, we know that all holomorphic Hamiltonian vector
field actions on $\mathfrak M_0$ lift to an action on the projective symplectic
resolution $\mathfrak M_{\vec\zeta}$, thus the resolution is equivariant.
\paragraph{}
In the following we will often specialize to the case of a singular Nakajima
quiver variety  $\mathcal X:=\mathfrak M_0$ and its resolution $\tilde{\mathcal
X}:=\mathfrak M_{\zeta_\bR,0}$, where $(\zeta_\bR,0)$ is generic. In the more
general case of a hyperK\"ahler cone that is not a Nakajima quiver variety, we
restrict to cones such that the  projective symplectic resolution exists, and
that it is
equivariant
with respect to the $\bC^\times\times G_H$ action. We will abuse notation by
using $\mathcal J_i$ and $R$ to denote the corresponding
actions on $\tilde{\mathcal X}$.
\paragraph{}
We are now ready to define our regularised \emph{superconformal
  index},
$\mathcal Z$:
\baa\label{Definition of superconformal index}
\mathcal Z(\tilde{\mathcal
X};\tau,y,Z):&=\sum_{p,q=0}^{2d_H}(-)^{p+q}\left(\frac
y\tau\right)^{p-d_H}\tr_{H^q(\tilde{\mathcal X},A^p(\tilde{\mathcal
X}))}\left(\tau^{R}\prod_iz_i^{\mathcal J_i}\right)\,.
\eaa
Note that $\mathcal Z$ is an analog of the Hirzebruch $\chi_{-y}$-genus of
$\tilde{\mathcal X}$.

\paragraph{}
For a hyperK\"ahler cone $\mathcal X$ with a $\bC^\times$-action such  that the
space of polynomial-valued holomorphic functions on $\mathcal X$ is
non-negatively graded under $\bC^\times$, the zeroeth graded component being
solely the constant functions, and the holomorphic symplectic form is
homogeneous with respect to this grading, Namikawa in
\cite{Namikawa:2013poisson} showed that there are only finitely many
non-isomorphic projective symplectic resolutions of $\mathcal X=\mathfrak M_0$.
We may ask, given two non-isomorphic equivariant projective symplectic
resolutions of $\mathcal X$, $\tilde{\mathcal X}$ and $\tilde{\mathcal X}'$, do
they have the same superconformal index? If this is the case, then the
index computed on $\tilde{\mathcal X}$ can be regarded as
an invariant of $\mathcal X$, and
our regularisation by working on the resolution makes sense.
We conjecture that this is indeed the case for all such hyperK\"ahler cones,
and will
write
 \be
 \mathcal Z\equiv\mathcal Z(\mathcal X)\,.
 \ee
We will present various pieces of evidence for this in the following.
In particular we will prove that this property holds in various limits
and specialisations of the index. We also perform some explicit
calculations to verify this property for quivers of low dimension.
\paragraph{}
First, we note that that any
choice of projective symplectic resolution of a Nakajima quiver variety will
give the same index if we set $\tau$ to 1. This is because of the following
theorem:
\begin{theorem}\label{Nowallcrossing} (A simple generalisation of 3.4 in
\cite{Proudfoot:2004hyperkahler})
If $\vec\zeta$ and $\vec\zeta'$ are both generic, then $\mathfrak
M_{\vec\zeta}$ and $\mathfrak M_{\vec\zeta'}$ are $G_H$-equivariant
diffeomorphic.
\end{theorem}
Essentially this theorem holds because provided we forget about the
$\mathbb{C}^\times$-grading by setting $\tau=1$ we are free to turn on
$\zeta_\bC$, and so can form a homotopically unique path through regular values
from one resolution
to the other. This ceases to be true if we turn on $\tau$, which grades with
respect to a preferred complex structure. To preserve the
$\mathbb{C}^\times$-grading we are required to set $\zeta_\bC=0$ and
typically the space of generic values on this slice of the parameter
space is disconnected by walls of codimension one.  So we are claiming that
there is no wall crossing for the superconformal index of a resolved Nakajima
quiver variety. We discuss this in more detail in Appendix
\ref{WallCrossingAppendix} with examples.

Two  further pieces of evidence for our conjecture are that two limits of the
superconformal index discussed in section \ref{ThePoincarePolynomial}, the
Poincar\'e polynomial and the Hilbert series, are known to be the same for
different resolutions. The Poincar\'e polynomial does not depend on the choice
of resolution, because all resolutions are  diffeomorphic. The Hilbert series
does not depend on the choice of resolution, because it is equal to the count
of holomorphic functions on the unresolved space, \cite{Kaledin:2003crepant}.

\subsection{Localisation}\label{Localisation}

Now that we have defined the index, it remains to compute it. We show that in
many cases the superconformal index can be computed exactly. We write $T$ for
the maximal torus of $\bC^\times\times G_H$, and will assume that $\mathcal X$
has a unique fixed point under $T$ and that $\tilde{\mathcal X}$ has finitely
many  fixed points under $T$. This is indeed the case for all Nakajima
quiver varieties that we investigate in this paper.
\paragraph{}
We will express $\mathcal Z$ as a sum of equivariant Euler
characters of equivariant sheaves. This means that we are able to use
localisation theorems from equivariant K-theory to exactly compute $\mathcal Z$
as a sum over fixed points. Writing $\chi_T$ for the $T$-equivariant Euler
character, it is defined for any holomorphic $T$-equivariant coherent sheaf
$\mathcal V$ as
\be
\chi_T(\mathcal
V):=\sum_{i=0}^{2d_H}(-)^i\,\text{ch}_T\,H^i(\tilde{\mathcal
X},\mathcal V)\,.
\ee
It is immediate from  equation (\ref{Definition of superconformal index}) that
\be\label{noteqeulercharacter}
\mathcal Z(\mathcal X;\tau,y,Z):=\sum_{p=0}^{2d_H}(-)^{p}\left(\frac
y\tau\right)^{p-d_H}\chi_T(A^p(\tilde{\mathcal X}))\,.
\ee
\paragraph{}
We pause to note the importance of the factor of $\tau^{d_H-p}$. If we had
not taken this factor out, then we would have been grading with respect to
$-\bM-2\bJ_3$. This is an action on the space of forms, but not an action on
the
base manifold and hence we cannot form an equivariant action on the sheaves
with this action. However, $R$ and $\mathcal J_i$ are actions on $\mathcal X$,
and hence we have written $\mathcal Z$ as a sum of equivariant Euler
characters.
\paragraph{}

The localisation theorem for equivariant K-theory is due to
\cite{Thomason:1992}, though a prelimary version is in \cite{Atiyah:1968index}.
This theorem tells us that the natural inclusion map, $\iota:\tilde{\mathcal
X}^T\to \tilde{\mathcal X}$, of the $T$-fixed points of $\tilde{\mathcal X}$
into $\tilde{\mathcal X}$ induces a homomorphism, $\iota_*$, that is an
isomorphism after localisation.

\begin{proposition} (from \cite{Nakajima:2005instanton})  Let $\mathcal V$ be a
$T$-equivariant locally free sheaf on $\tilde{\mathcal X}$. Then we have that
\be\label{aneqeulercharacter}
\chi_T(\mathcal
V)=\sum_{x\in\tilde{\mathcal X}^T}\iota^*_x(\mathcal
V)\text{\emph{PE}}[\text{\emph{ch}}_T(T^*_x\tilde{\mathcal X};\tau,Z)]\,.
\ee
Here $\tilde{\mathcal X}^T$ is the space of $T$-fixed points of
$\tilde{\mathcal X}$, $\iota_x:\{x\}\hookrightarrow\tilde{\mathcal X}$ is the
inclusion of the
fixed point $x$ in $\tilde{\mathcal X}$, and $\chi$ is the $\bC^\times\times
G_H$-equivariant Euler character.
\end{proposition}

PE is the plethystic exponential. It is defined as
\be\label{PEdef}
\text{PE}[f(t_1,\dots,t_n)]:=\text{exp}\left(\sum_{r=1}^\infty\frac{f(t_1^r,\dots,t_n^r)}r\right)\,.
\ee
The plethystic exponential of a polynomial is as follows
\be
\text{PE}\left[\sum_i t_i-\sum_j
s_j\right]=\frac{\prod_j(1-s_j)}{\prod_i(1-t_i)}\,,
\ee
where the $t_i$ and $s_j$ are monomials.
\paragraph{}
We can now compute our superconformal index using localisation.
\baa\label{SuperconformalIndexDerivation}
\mathcal Z(\mathcal X)&=\sum_{p=0}^{2d_H}(-)^p\left(\frac
y\tau\right)^{p-d_H}\chi(A^p(\tilde{\mathcal X});\tau,Z)\\
&=\sum_{p=0}^{2d_H}(-)^p\left(\frac
y\tau\right)^{p-d_H}\sum_{x\in\tilde{\mathcal
X}^T}\text{ch}_T(\Lambda^p(T^*_x\tilde{\mathcal
X});\tau,Z)\text{PE}[\text{ch}_T(T^*_x\tilde{\mathcal X};\tau,Z)]\\
&=\left(\frac\tau y\right)^{d_H}\sum_{x\in\tilde{\mathcal
X}^T}\text{PE}\left[\left(1-\frac
y\tau\right)\text{ch}_T(T^*_x\tilde{\mathcal X};\tau,Z)\right]\,.
\eaa
We introduce the following notation for the contribution at each fixed
point,
\baa
\mathcal Z&\equiv\left(\frac\tau y\right)^{d_H}\sum_{x\in\tilde{\mathcal
X}^T}\mathcal Z_x\\
&\equiv\left(\frac\tau y\right)^{d_H}\sum_{x\in\tilde{\mathcal
X}^T}\text{PE}\left[\left(1-\frac y\tau\right)\sum_{\alpha\in\mathcal
J_x}m_\alpha(\tau,Z)\right]\,,
\eaa
where $\alpha=(\alpha_0,\gamma)$, $\alpha_0\in\bZ$ and $\gamma$ is a weight
of $G_H$; $m_\alpha(\tau,Z)=\tau^{\alpha_0}Z^\gamma$ is a monomial; and
$\mathcal J_x$ is the collection of $T$-weights of the module
$T^*_x(\tilde{\mathcal X})$.

\paragraph{}
An alternative procedure to evaluating these quantities is via the use of the
Jeffrey-Kirwan localisation theorem, \cite{Jeffrey:1995localization}. This
theorem is for the evaluation of the integral of a form over a symplectic
quotient. In the case that $\tilde{\mathcal X}$ is a hyperK\"ahler quotient, we
can reduce the evaluation of $\mathcal Z(\mathcal X)$ to this problem via the
use of Grothendieck-Riemann-Roch theorem. We can then use the different choices
of chambers in the Jeffrey-Kirwan residue, corresponding to the different
choices of resolutions, to explicitly see that the superconformal index does
not depend on the choice of resolution.

\section{Properties of the superconformal index}

In order for $\mathcal{Z}(\mathcal{X})$ defined above to be
a superconformal index of $\mathfrak{osp}(4^*|4)$ representations, it is
necessary that it coincides with the form predicted in (\ref{general}). 
Notably, there must be a $\bZ[Z^{\pm1}]$-expansion in 1/2- and 1/4-BPS
$\mathfrak{su}(1|2)$ characters, and the coefficients of all 1/2-BPS
representations are positive integers with no $Z$-dependence\footnote{Note that
there is a condition on the coefficients of the 1/4-BPS states $S(n+1,n)$. We
do not investigate this condition in this paper.}. We shall show that this is true, at least for all examples we investigate.
 \paragraph{}

 We use the result in appendix \ref{su21characterexpansion}, this tells us that if
$\mathcal Z$ obeys four equations then it obeys the necessary property
to be a $\mathfrak{osp}(4^*|4)$ superconformal index listed above. There is a further positivity
condition imposed on the coefficient of the semishort representations
$SS(j/2,d_H-1,d_H-1)$. We show this is indeed true by investigating what we
call the Hilbert series limit of the superconformal index. 
\paragraph{}
The four
equations are
\baa\label{3propertya4}
\mathcal Z(\tau,y,Z)&=\mathcal Z(\tau,1/y,Z)\,.
\eaa

Writing
\baa
\mathcal
Z(\tau,y,Z)&=\sum_{a=0}^\infty\sum_{b=-d_H}^{d_H}\alpha_{a,b}(Z)\tau^ay^b=\sum_{a=0}^\infty\sum_{b=-m}^m\alpha_{a,b}(Z)\tau^{a+b}\left(\frac
y\tau\right)^b\,,\\
\eaa
we have
\begin{align}
\alpha_{a,b}&=0\text{ for }a<|b|\,,\label{3propertya6}\\
\mathcal Z(\tau,\tau,Z)&\in\bZ_{\geqslant0}\,,\text{ so that
}\frac{d}{d\tau}\mathcal Z(\tau,\tau,Z)=\frac{d}{dz_i}\mathcal
Z(\tau,\tau,Z)=0\,,\label{3propertya7}\text{ and }\\
\lim_{\substack{\tau\to0\\y/\tau\text{ finite}}}&\mathcal
Z(\tau,y,Z)=\sum_{a=0}^{d_H}\alpha_{a,-a}(Z)\left(\frac\tau
y\right)^a\in\bZ_{\geqslant0}\left[\frac \tau y\right]\label{3propertya8}\,.
\end{align}

Note that equation (\ref{3propertya4})  is manifest\footnote{If we taken $\tau$ to be
the fugacity for the scaling symmetry, $R$, this would be equivalent to using a
fugacity $\tilde y:=y\tau$ to count the $p$-grading of a $(p,q)$-form. We would
have had less trouble using equivariant localisation theorems, but at the
cost of losing the manifest $y\mapsto1/y$ symmetry.}. This follows because the
fugacity $y$ corresponds to the Cartan generator of the $SU(2)$ Lefschetz
action, whose raising operator is wedging with $\omega_\bC$. The resulting
$y\mapsto1/y$ symmetry is known as  Serre duality. Equation (\ref{3propertya7}) is true for all examples where the resolution has isolated fixed points and equation (\ref{3propertya8}) is true for all Nakajima quiver varieties.

\paragraph{}
We show that equation (\ref{3propertya6}) holds in all examples we investigate.
In all examples we investigate we find that the $y\mapsto1/y$ symmetry is
preserved at the level of the fixed point contributions to $\mathcal Z$. We
shall use to show that in a Taylor expansion in powers of $\tau$, for all
monomials of the form $\tau^ay^b$ we have $a\geqslant|b|$.

\paragraph{}
In the superconformal index, the contribution from each fixed points has a
factor of $1-\tau/y$ in the plethystic exponential.  Assuming a
$y\mapsto 1/y$ symmetry at each fixed point, there must also be
a factor of $1-y\tau$. Furthermore, since the solution is a finite polynomial
in $y$, we expect to be able to write the superconformal index as
\be\label{factoroutyintwoways}
\mathcal Z=\sum_{x\in\tilde{\mathcal X}^T}\text{PE}\left[p_x(\tau,Z)(1-\tau
y)(1-\tau/y)\right]\,,
\ee
where $p_x$ is a Laurent polynomial in $\tau$ and $Z$ with positive integer
coefficients, $p_x(\tau,Z)\in\bZ_{\geqslant0}[\tau^{\pm1},Z^{\pm1}]$. One can
then directly calculate that equation (\ref{3propertya6}) does indeed hold for
every fixed point.

\paragraph{}
Note that in order for equation (\ref{factoroutyintwoways}), it is
necessary that, for $x$ a $T$-fixed point of $\tilde{\mathcal X}$,
\be\label{FixedPointPolynomial}
\text{ch}_T(T^*_x(\tilde{\mathcal
X});\tau,Z)=p_x(\tau^{-1},Z^{-1})+\tau^2p_x(\tau,Z)\,.
\ee
\paragraph{}
Equation (\ref{3propertya7}), that the superconformal index at $\tau=y$ is a
positive integer, is an immediate consequence of equation
(\ref{SuperconformalIndexDerivation}).
\paragraph{}

In order to conclude that $\mathcal Z(\mathcal X)$ is in the form of a
$\mathfrak{osp}(4^*|4)$ superconformal index it is necessary that equation
(\ref{3propertya8}) holds, namely
\be
\lim_{\substack{\tau\to0\\y/\tau\text{ finite}}}\mathcal Z_{\mathcal
X}(\tau,y,Z)\in\bZ_{\geqslant0}\left[\frac y \tau\right]\,.
\ee
We shall show this in the next subsection for all Nakajima quiver varieties. It
follows from the fact that the $\tau\to0$ limit with $y/\tau$ fixed of the
superconformal index is the Hirzebruch $\chi_{-y}$-genus of the
$\bC^\times$-fixed
point submanifold\footnote{It is not a superconformal index, as this space is
not the resolution of a cone.}.  From \cite{Nakajima:2001quiver}, it is known
that this
submanifold is compact, and hence the superconformal index is the Poincar\'e
polynomial, and moreover it is known the space's odd homology vanishes. Hence
the superconformal index is a positive polynomial, the fact that $\mathcal
Z_{\mathcal X}(\tau,\tau,Z)\in\bZ_{\geqslant0}$ means that the $\tau\to0$ limit
cannot depend on $Z$.

\subsection{Limits of the  superconformal index}\label{ThePoincarePolynomial}

In this subsection, we shall consider three limits of the superconformal index
which exhibit interesting behaviour.  In particular, we consider limits where
the superconformal index  reduces to the the Poincar\'e polynomial of
$\mathcal{X}$ and to its Hilbert series. Finally, we show that if one
hyperK\"{a}hler cone $\mathcal{Y}$ is contained inside another $\mathcal{X}$ as
a fixed point subspace of a triholomorphic isometry  then there is a limit in
which the superconformal index of $\mathcal{X}$ reduces to that of
$\mathcal{Y}$.
\paragraph{}
We start by considering  the generating series for Borel-Moore homology. For
$M$ a manifold (possibly non-compact), the Borel-Moore homology of $M$  is
defined as the relative singular homology of the one point compactification of
$M$, $\bar M$, with respect to the point at infinity, and so for compact
manifolds the Borel-Moore homology is identical to the singular homology.
\be
H^{BM}_*(M):=H_*(\bar M,\{\infty\})\,.
\ee

Pick a $\bC^\times$-action on $\tilde{\mathcal X}$ defined by some
\be
\lambda:\bC^\times\hookrightarrow \bC^\times\times
T_H=T\cong(\bC^\times)^{r+1}\,.
\ee
We assume that this is a \emph{generic} action, this means that it has isolated
fixed points. We further assume that $\lim_{t_1\to0}\lambda(t_1)=\infty$. We
have defined $r:=\text{rk}(G_H)$.
\paragraph{}
Theorem 3.7 (3) and (4) of \cite{Nakajima:03} easily lift to any Nakajima
quiver variety with isolated fixed points. It tells us that the homology
vanishes at odd degree; is freely generated at even degree; and that each fixed
point contributes a single generator, whose homology degree is given by the
dimension of the (+)-attracting set at that point. That is, for
$x\in\tilde{\mathcal X}^T$ a fixed point, the (+)-attracting set is
\be
S_x=\{p\in\tilde{\mathcal X}\big|\lim_{t\to0}\lambda(t)\cdot p=x\}\,.
\ee
We then define the Poincar\'e polynomial as the generating function of
equivariant Borel-Moore homology:
\baa
P_{\tilde{\mathcal
X}}(q)&:=\sum_{i=0}^{2d_H}\dim\big(H_{2i}^{BM}(\tilde{\mathcal X})\big)q^i\\
&=\sum_{x\in\tilde{\mathcal X}^T} q^{\dim_\bC S_x}\,.
\eaa
\paragraph{}
 We now show that the contribution to the superconformal index
  at each fixed point of $\tilde{\mathcal X}$, contains information
  about the dimension of the (+)-attracting set. A generic choice of
$\lambda$ is given by
\be
\lambda(t)=(t^{m},t^{n_1},\dots,t^{n_r})
\ee
for some
\be
0> n_1>\dots>n_r\gg m\,.
\ee
We write our superconformal index as a function of the fugacities $\tilde
y,\tau,Z$, where $\tilde y=y/\tau$. Then one makes the following
replacements for the fugacities appearing
in the formula (\ref{SuperconformalIndexDerivation})
for $\mathcal{Z}_{\mathcal{X}}$;
\be
\tau\mapsto s^m\,,\quad z_i\mapsto s^{n_i}\,,
\ee
for $s$ a non-zero complex number.
Finally, one takes the limit $s\to0$. Because the powers of $s$
appearing in the numerator and denominator of each factor of
$\mathcal{Z}_{x}$ agree, the limit of the index is a
Laurent polynomial in $\tilde y$ with positive integer coefficients.
\paragraph{}
 For a particular fixed point, $x\in\tilde{\mathcal X}^T$, one has a product of
the form
 \baa
\mathcal Z_x(\tilde y,s)&=\tilde y^{-d_H}\prod_{\alpha\in\mathcal
J_x}\frac{1-\tilde
ys^{\alpha_0m+\sum_i\gamma_in_i}}{1-s^{\alpha_0m+\sum_i\gamma_in_i}}\\
&=:\tilde y^{-d_H}\prod_{\alpha\in\mathcal J_x}\frac{1-\tilde
ys^{\ell_\alpha}}{1-s^{\ell_\alpha}}\,,
\eaa
where we have defined $\ell_\alpha:=\alpha_0m+\sum_i\gamma_in_i$. Note that
the $\bC^\times$-action being generic necessarily means that $\ell_\alpha\neq0$
for all $\alpha\in\mathcal J_x$ for all $x\in\tilde{\mathcal X}^T$. We then
take the limit $s\to0$ and obtain
\be
\lim_{s\to0}\mathcal Z_x=\tilde y^{-d_H+|\{\alpha:\ell_\alpha<0\}|}
\ee
Since the sign of $\ell_\alpha$ tells us about whether the tangent
direction $\alpha$ at the fixed point is an attracting or repelling one, we
have that $|\{\alpha|\ell_\alpha<0\}|=\dim_\bC S_x$. Thus, the Poincar\'e
polynomial is given by
\baa
P_{\tilde{\mathcal X}}(\tilde y)=\tilde y^{d_H}\lim_{s\to0}\mathcal
Z_{\mathcal X}\left(\tilde y, \tau=s^{m}, z_{i}=s^{m_{i}} \right)\,.
\eaa
\paragraph{}
Since the superconformal index is a grading under $\mathfrak{su}(1|2)$ and
global symmetries $G_H$, it has a character expansion as
\be
\mathcal Z=\sum_{R_1,R_2}
\chi_{\mathfrak{su}(1|2)}(R_1;\tau,y)\chi_{G_H}(R_2;Z)\,,
\ee
where the sum is over finite dimensional irreducible representations of
$\mathfrak{su}(1|2)$ and $G_H$; and $\chi_G(R;W)$ is the character of the
representation $R$ of $G$, with fugacities $w_1,\dots,w_{\text{rk}(G)}$.
\paragraph{}
In equation (\ref{SU21characters}), we find the
characters of all finite dimensional irreducible representations of
$\mathfrak{su}(1|2)$. Importantly, if we set $\tau=y$, then we see that the
only non-zero contributions to the index are from the 1/2-BPS short
multiplets $S(m,m)$, which each contribute unity to the
superconformal index. If we look in the formula
(\ref{SuperconformalIndexDerivation}) for the superconformal index, we see that
we get contribution of 1 for each fixed point, so we may identify each fixed
point with a 1/2-BPS multiplet.
\paragraph{}
If we keep $\tau/y$ fixed while sending $\tau$  as well as the fugacities $Z$
to zero we compute the Poincar\'e polynomial with grading by homological
degree. Since the 1/2-BPS states are necessarily invariant under $G_H$ (see
\cite{Aharony:1997light}), one can see that the only terms that survive in
$\mathcal Z$ are the 1/2-BPS multiplets. Their contribution is $(\tau/y)^m$,
where
$m$ is the highest power of $y$ that was in the $\mathfrak{su}(1|2)$ character.
\paragraph{}
This means that if $P=\sum_{n=0}^{d_H}b_{2n}q^n$ is the Poincar\'e
polynomial\footnote{The reason that this sum goes up to $d_H$ and not to
$2d_H$ as one might naively expect, is because for Nakajima quiver varieties
$\mathcal X$ is homotopic to one of its Lagrangian subvarieties, for example
see \cite{Ginzburg:2009lectures}.} of $\tilde{\mathcal X}$, then we can
reconstruct the 1/2-BPS state spectrum as
\be\label{1/2-BPS states}
\mathcal H_{\text{1/2-BPS}}=\bigoplus_{n=0}^{d_H}b_{2(d_H-n)}S(n,n)\,.
\ee

Note that this is nothing more than the statement that the 1/2-BPS
multiplets of the superconformal algebra are in one-to-one
correspondence with
the compactly supported cohomology classes of $\tilde{\mathcal X}$, and
Poincar\'e duality relates this to the Borel-Moore homology. In the
case where $\mathcal{X}$ is the moduli space of Yang-Mills instantons,
this correspondence was anticipated in \cite{Aharony:1997light}.
\paragraph{}
Equation (\ref{1/2-BPS states}) clearly shows that the multiplicities
of 1/2-BPS states are
non-negative integers independent of the flavour fugacities. Together
with the earlier results in this section, this means that
$\mathcal Z$ satisfies all the criteria imposed by the condition that it is a
superconformal index of an $\mathfrak{osp}(4^*|4)$ representation.
\paragraph{}
In the works \cite{Hausel:05,Hausel:08}, Hausel gives the generating function
for the Poincar\'e polynomial of any Nakajima quiver variety. Assuming that the
variety has isolated fixed points, we can then use this to give the full
1/2-BPS spectrum of the theory.
\paragraph{}
While further results such as \cite{Mozgovoy:2006}, mean that for $A$-type
quivers, as well as the Coulomb
branch of $DE$-type quivers, the  1/2-BPS states highest states are given by
the fusion product of fundamental Kirillov-Reshitikhin modules of $ADE$-type,
and can be given by Hatayama's fermionic form \cite{Hatayama:1998}.

\paragraph{}

Now we discuss the Hilbert series limit of the superconformal index. In the
works
\cite{Cremonesi:2014monopole,Cremonesi:2014coulomb,Hanany:14,Hanany:14glue,Cremonesi:2014sicilian},
the Hilbert series, HS, is the count of polynomial valued holomorphic functions
on $\mathcal X$ graded by global symmetries and the $\bC^\times$-action. Note
that for any variety $Y$,  $\Gamma(Y,\mathcal O_Y)$ is defined as the space of
polynomial valued global holomorphic functions on $Y$. The Hilbert series is
\be
\text{HS}(\mathcal X)=\tr_{\Gamma(\mathcal X,\mathcal O_{\mathcal
X})}\left(\tau^R\prod_iz_i^{\mathcal J_i}\right)\,.
\ee

One can easily see from the definition of $\mathcal Z$ that the coefficient of
$y^{d_H}$ divided by $\tau^{d_H}$ is $\chi(\mathcal O_{\tilde{\mathcal X}})$,
the
equivariant Euler character of the structure sheaf. Explicitly,
\be
\lim_{y\to\infty}y^{-d_H}\tau^{-d_H}\mathcal Z_{\mathcal
X}=\sum_{q=0}^{2d_H}(-)^q\tr_{H^{0,q}(\tilde{\mathcal
X})}\left(\tau^R\prod_iz_i^{\mathcal J_i}\right)\,.
\ee
 For Nakajima quiver varieties, we have that the space of holomorphic functions
on $\tilde{\mathcal X}$ and $\mathcal X$ are isomorphic as graded Poisson
algebras. Namely,
\be\label{coordring1}
\pi^*:\Gamma(\mathcal X,\mathcal O_{\mathcal X})\to\Gamma(\tilde{\mathcal
X},\mathcal O_{\tilde{\mathcal X}})
\ee
is an isomorphism of graded Poisson algebras, see \cite{Kaledin:2003crepant}.
\paragraph{}
 We further have (from \cite{Ginzburg:2009lectures}) that for all $q>0$,
\be\label{coordring2}
H^q(\tilde{\mathcal X},\mathcal O_{\tilde{\mathcal X}})=0\,.
\ee

This means that for Nakajima quiver varieties the coefficient of $y^d$ in
$\mathcal Z$ divided by $\tau^d$ is the Hilbert series\footnote{Note because of
the $y\mapsto1/y$ symmetry, one can swap the limit of $y$ to $\infty$ for a
limit of $y$ to $0$, at the cost of multiplying by $y^{d_H}$ instead of
$y^{-d_H}$.}
\be
\text{HS}(\mathcal X)=\lim_{y\to\infty}y^{-d_H}\tau^{-d_H}\mathcal Z_{\mathcal
X}\,.
\ee
Like the limit corresponding to the Poincare polynomial, the Hilbert
series limit also yields a precise counting of certain superconformal
multiplets. In particular, a holomorphic function of charge
$r\in \mathbb{N}_{0}$ under
the $\mathbb{C}^{\times}$-action generated by $R$ corresponds to a (semi)-short
multiplet,
\bea
S(d_{{H}},d_{{H}}) && \text{for}\,\, r=0 \nonumber \,,\\
S(d_{{H}},d_{{H}}-1) && \text{for}\,\, r=1 \nonumber\,, \\
SS(R/2-1,d_{{H}}-1,d_{{H}}-1) && \text{for}\,\, r\geq 2\,.
\nonumber
\eea
This means that
\be
\text{HS}(\mathcal X)=N^{(d_H,d_H)}+N^{(d_H,d_H-1)}\tau+\sum_{r=2}^\infty
N^{(r/2-1,d_H-1,d_H-1)}\tau^{r}\,.
\ee
\paragraph{}
Finally, we discuss the limit of the superconformal index that gives the
superconformal index of a fixed point subspace. Let $\mathcal X$ be a
hyperK\"ahler cone, $\tilde{\mathcal X}$ its projective symplectic resolution
with isolated fixed points under $T=\bC^\times\times T_H$. If $\tilde{\mathcal
N}\subset \tilde{\mathcal X}$ is the fixed point subspace (generally not
connected) under a closed Lie subgroup of isometries $T'\subset T_H$, then we
have that $\tilde{\mathcal N }$ will be the projective symplectic resolution of
a disjoint union of hyperK\"ahler cones $\mathcal N$. This follows from the
fact that the resolution is $T_H$-equivariant.
\paragraph{}
The superconformal index of $\mathcal N$ will be graded by $\bC^\times\times
T_H/T'$. Suppose without loss of generality that $T'=U(1)$ (without loss of
generality as we can do this rk\,$T'$ times). We write $\tilde z_i$ for
$i=1,\dots,\text{rk}\,T_H-1$ as the fugacities of $T_H/T'$. The inclusion of
$T'$ into $T_H$ defines a relabelling of fugacities $z_i\mapsto s^{f_i}\tilde
z_{i'(i)}$, where $f_i\in\bZ$ and $s$ is the fugacity for $T'=U(1)$. The
superconformal index of $\mathcal N$ will be the superconformal index of
$\mathcal X$, but with the all cotangent directions at a fixed point in
$\tilde{\mathcal X}$ that have any charge under $s$ thrown away. One can
achieve this by sending $s\to 0$:
\be
\mathcal Z_{\mathcal N}(\tau,y;\tilde Z)=\lim_{s\to0}\mathcal Z_{\mathcal
X}(\tau,y;s^{f_1}\tilde z_{i'(1)},\dots,s^{f_{\text{rk}(T_H)}}\tilde
z_{i'(\text{rk}(T_H))})\,.
\ee

A similar limit exists where we send $\tau\to0$, restricting to the fixed point
submanifold of the $\bC^\times$-action, $\tilde{\mathcal X}^{\bC^\times}$. This
necessarily breaks the hyperK\"ahler structure, but the superconformal index,
$\mathcal Z(\tilde{\mathcal X}^{\bC^\times})$, is still defined. The space
$\tilde{\mathcal X}^{\bC^\times}$ is projective and connected (lemma 7.3.3 and
proposition 7.3.4 of \cite{Nakajima:2001quiver}) for all Nakajima quiver
varieties. This means that the de Rham cohomology and the Borel-Moore homology
coincide. One can easily compute that
\baa
\mathcal Z(\tilde{\mathcal
X}^{\bC^\times})&=\left(\frac{\tau}{y}\right)^{d_H-|\{\alpha\in\mathcal
J_x|\alpha_0<0\}|} \sum_{x\in\tilde{\mathcal
X}^{\bC^\times}}\prod_{\substack{(\alpha_0,\gamma)\in\mathcal
J_x\\\alpha_0=0}}\frac{1-\frac{y}{\tau}Z^\gamma}{1-Z^\gamma}\\
&=\sum_r\tr_{H^r(\tilde{\mathcal
X}^{\bC^\times})}\left((-)^ry^\bN\prod_iz_i^{\mathcal J_i}\right)\,.
\eaa
It is known that the odd cohomology vanishes. So, we have that namely that
$\mathcal Z(\tilde{\mathcal
X}^{\bC^\times})\in\bZ_{\geqslant0}((Z))[\tau/y^{\pm1}]$. However, we also have
that $\mathcal Z_{\mathcal X}(\tau,\tau)\in\bZ_{\geqslant0}$. From this we may
conclude that
\be
\lim_{\substack{\tau\to0\\\frac{y}{\tau}\text{ fixed}}}\mathcal
Z(\tilde{\mathcal X}^{\bC^\times})\in\bZ_{\geqslant0}\left[\frac\tau
y\right]\,.
\ee
This confirms that equation (\ref{3propertya8}) does indeed hold, and hence the
superconformal index does obey the necessary properties to be a superconformal
index of $\mathfrak{osp}(4^*|4)$.

\paragraph{}
We consider a simple example to illustrate each of the different
limits. Take $\mathbb C^4$, with coordinate ring $\mathbb C[X_1,\tilde
X_1,X_2,\tilde X_2]$.
$X_i$ is
charged\footnote{One should think of this as the charge of the operator given
by multiplication by $X_i$, and similarly for the other variables.} as $\tau
s_i$ and $\tilde X_i$ is charged as $\tau /s_i$ for i=1,2. $\tau /s_i$ and
$\tau s_i$ are fugacities for $\mathbb C^\times_{i,1}\times\mathbb
C^\times_{i,2}$ rotating the target space $\mathbb C^2_i=\mathbb
C_{i,1}\times\mathbb C_{i,2}$. The diagonal subgroup of $\mathbb
C^\times_{1,1}\times\mathbb C^\times_{1,2}$ is counted with the same fugacity
as the diagonal subgroup of $\mathbb C^\times_{2,1}\times\mathbb
C^\times_{2,2}$, this is the $\bC^\times$ generated by $R$. $dX_i$ and $d\tilde
X_i$ are charged as $y s_i$ and $y /s_i$ respectively, where the fugacity $y$
is for a  $\mathbb C^\times_y$ that rotates the cotangent fibres. The
superconformal index is,
\be
\mathcal Z_{\bC^4}(\tau,y;s_1,s_2)=\left(\frac {\tau}
y\right)^2\prod_{i=1}^2\frac{(1-y s_i)(1-y
/s_i)}{(1-\tau s_i)(1-\tau /s_i)}\,.
\ee
The Hilbert series is the coefficient of the highest power of $y$ divided by
$\tau^2$,
\be
\text{HS}(\bC^4)=\prod_{i=1}^2\frac{1}{(1-\tau s_i)(1-\tau /s_i)}\,.
\ee
The Borel-Moore homology of $\bC^4$ has only one generator, the fundamental
class
$[\bC^4]$. Hence
\be
H^{\text{BM}}_i(\bC^4)=\begin{cases}\bZ\,\text{ if }i=8\,,\\0\,\text{
otherwise.}\end{cases}
\ee
To form the Poincar\'e polynomial we rewrite $\mathcal Z$ in terms of $\tilde
y$, $\tau\mapsto s^{-5}$, $s_1\mapsto s^{-2}$, and $s_2\mapsto s^{-1}$. This
gives
\be
\mathcal Z_{\bC^4}(s^{-5},s^{-5}\tilde y;s^{-2},s^{-1})=\tilde
y^{-2}\frac{(1-\tilde y s^{-7})(1-\tilde y s^{-3})(1-\tilde y s^{-6})(1-\tilde
y s^{-4})}{(1-s^{-7})(1- s^{-3})(1- s^{-6})(1- s^{-4})}\,.
\ee
From this we see
\be
P_{\bC^4}(\tilde y)=\tilde y^4=\tilde y^2\lim_{s\to0}\mathcal
Z_{\bC^4}(s^{-5},s^{-5}\tilde y;s^{-2},s^{-1})\,.
\ee
We consider restricting to the hyperK\"ahler submanifold invariant under the
subgroup $\{(x,1/x)|x\in\mathbb C^\times\}\subset \mathbb
C^\times_{1,1}\times\mathbb C^\times_{1,2}$, namely $\mathbb
C_{2,1}\times\mathbb C_{2,2}$. We do this by discarding all generators with
non-zero power of $s_1$. To take this limit, we take the limit $s_1\to0$ in the
index. This gives
\be
\mathcal Z_{\bC^2}(\tau,y;s_2)=\frac \tau y\frac{(1-y s_2)(1-y /s_2)}{(1-\tau
s_2)(1-\tau
/s_2)}\,.
\ee
We then get the Hilbert series by taking the highest power of $y$ divided by
$\tau$,
\be
\text{HS}(\bC^2)=\frac{1}{(1-\tau s_2)(1-\tau /s_2)}\,,
\ee
while the Poincar\'e polynomial is
\be
P_{\bC^2}(\tilde y)=\tilde y^2=\tilde y\lim_{s\to 0}\mathcal
Z_{\bC^4}(s^{-5},s^{-5}\tilde y;s^{-2})\,.
\ee

\section{Examples}

\subsection{Instanton moduli space}\label{ADHMExample}

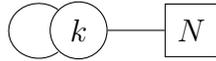
\begin{figure}[h]
\centering
\begin{tikzpicture}[]
  \node[minimum size=0.7cm,draw, circle](majambe) at (-5,0.0){$k$};
\node[minimum size=0.7cm,draw](monbras) at (-3.5,0.0){$N$};
\draw[-](monbras.west)--(majambe.east);
\draw (majambe.north west) arc  [x radius=0.4, y radius=0.365, start angle=50,
end angle=310];
\end{tikzpicture}
\caption{The ADHM quiver.}\label{ADHMquiver}
\end{figure}

The Nakajima quiver variety associated to the ADHM quiver, in figure
\ref{ADHMquiver}, has known fixed points, with the associated tangent space at
the fixed point calculated in \cite{Nakajima:1999lectures,Nekrasov:02}. The
global symmetry is
\be
G_H=SU(N)\times SU(2)\,,
\ee
where the $SU(N)$ is the flavour symmetry associated to the box  and the
$SU(2)$ is a flavour symmetry associated to the adjoint of $U(k)$. The fields
are $X,\tilde X\in$End$(\bC^k)$, $Q\in$Hom$(\bC^N,\bC^k)$ and $\tilde
Q\in$Hom$(\bC^k,\bC^N)$. The moment map  equations defining the quiver are
known as the ADHM equation, \cite{Atiyah:1994construction}, and are
\baa
\mu_\bR&=[X,X^\dagger]+[\tilde X,\tilde X^\dagger]+QQ^\dagger-\tilde
Q^\dagger\tilde Q\,,\\
\mu_\bC&=[X,\tilde X]+Q\tilde Q\,.
\eaa
 We use fugacities $z_1,\dots,z_N$ for $SU(N)$ and
$x$ for $SU(2)$.
\paragraph{} The ADHM quiver has known fixed points under the action
of the Cartan torus of $G_{H}\times \mathbb{C}^\times$,
with the associated character of tangent space at
the fixed point calculated in \cite{Nakajima:1999lectures,Nekrasov:02}.
The fixed points are given by $N$-coloured Young tableaux of $k$. We define the
functions of a box, $s=(a,b)$, at row $a$ and column $b$ in the $i^\text{th}$
partition $Y_i$ of a coloured Young tableau $\vec Y$
\be
f_{ij}(s):=-a_i(s)-l_j(s)-1\,,\quad g_{ij}(s):=-a_i(s)+l_j(s)\,,
\ee
where $a_i(s):=Y_{ia}-b$ the arm length and $l_j(s):=(Y_j^\vee)_b-a$ the leg
length relative to $Y_j$. We can write the superconformal index of the ADHM
quiver as
\baa\label{Nekpartfn}
\mathcal
Z_{k,N}&=\sum_{\substack{\{ Y_i\}\\\sum_i|Y_i|=k}}\prod_{i,j=1}^N\prod_{s\in
Y_i}\text{PE}\left[\tau^{g_{ij}(s)-1}x^{f_{ij}(s)}\frac{z_i}{z_j}(1-\tau/y)(1-\tau
y)\right]\,,
\eaa
where $\substack{\{ Y_i\}\\\sum_i|Y_i|=k}$ is the sum over the $N$-coloured
Young tableaux corresponding to all $N$-coloured
partitions of $k$.
\paragraph{}
The superconformal index is therefore equal to the
$k$ instanton contribution to the Nekrasov partition function of
$\mathcal N=1^*$ five dimensional
supersymmetric Yang-Mills theory with gauge group
$SU(N)$ compactified to four dimensions on a circle.
This agrees with an earlier proposal for a
superconformal index of the ADHM moduli space quantum mechanics \cite{Kim:11}.
\paragraph{}
The dictionary between the parameters of the superconformal index and
those of the Nekrasov partition  is as follows: the parameters $\tau$ and $x$
are related to the deformation parameters for the $\Omega$-background via
\be
\tau=e^{\frac{\epsilon_1+\epsilon_2}{2}}\,,\quad
x=e^{\frac{\epsilon_1-\epsilon_2}{2}}\,;
\ee
if $m$ is the mass of the adjoint hypermultiplet, then
\be
y=e^m\,;
\ee
if $a_i$ for $i=1,\dots,N$ are the Coulomb branch parameters, then we have
\be
z_i=e^{a_i}\,.
\ee
 Here the five dimensional parameters are measured in
units of the radius of the compactification circle.

The Poincar\'e polynomial limit of the superconformal index reproduces the
result
\be
P(\tilde y)=\sum_{\substack{\{Y_i\}\\\sum_i|Y_i|=k}}\prod_{i=1}^N\tilde
y^{2N|Y_i|-2i\ell(Y_i)}\,.
\ee

\subsection{$\bC^2/\bZ_n$}\label{C2modZn}

In this subsection we work through an example where the orbifold cohomology of
\cite{Chen:2004} is the same as the cohomology of the symplectic resolution.

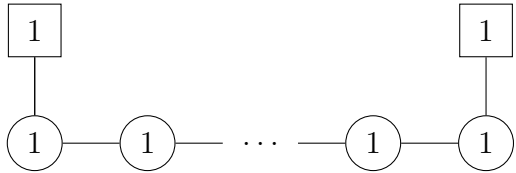
\begin{figure}[h]
\centering
\begin{tikzpicture}[]
\node[minimum size=0.6cm,draw, circle](lin1) at (0,0){$1$};
\node[minimum size=0.6cm,draw, circle](lin2) at (1.5,0){$1$};
\node at (3,0) {$\ldots$};
\node[minimum size=0.6cm,draw, circle](lin3) at (4.5,0){$1$};
\node[minimum size=0.6cm,draw, circle](lin4) at (6,0){$1$};
\node[minimum size=0.7cm,draw](flav1) at (0,1.5){$1$};
\node[minimum size=0.7cm,draw](flav2) at (6,1.5){$1$};
\draw[-](lin1.east)--(lin2.west);
\draw[-](lin2.east)--(2.5,0);
\draw[-](3.5,0)--(lin3.west);
\draw[-](lin3.east)--(lin4.west);
\draw[-](lin1.north)--(flav1.south);
\draw[-](lin1.north)--(flav1.south);
\draw[-](lin4.north)--(flav2.south);
\end{tikzpicture}
\caption{The quiver whose corresponding unresolved variety is $\bC^2/\bZ_n$.
There are $n-1$ nodes, $k=(1^{n-1})$ and $N=(1,0^{n-3},0)$. When $n=2$,
$N=(2)$, and the quiver is known as $T(SU(2))$.}\label{C2Znquiver}
\end{figure}

\paragraph{}
If $\mathcal X$ is a hyperK\"ahler orbifold, then  the cohomological
hyperK\"ahler resolution conjecture states that if
$\tilde{\mathcal X}$ is a hyperK\"ahler resolution of $\mathcal X$, then the
cohomology on $\tilde{\mathcal X}$ is the orbifold cohomology on $\mathcal X$.
See conjecture 6.3 of \cite{Ruan:2000stringy} for the first statement of this
conjecture, and \cite{Ruan:2001cohomology} for a slightly more sophisticated
wording of it. The orbifold cohomology was first defined in \cite{Chen:2004},
the key point is that it depends solely on $\mathcal X$, and so is independent
of the choice of resolution. It is known to be true for the following example
of $\bC^2/\bZ_n$. Due to how the orbifold cohomology is constructed, when
$\mathcal X$ is a hyperK\"ahler orbifold, the cohomology of $\tilde{\mathcal
X}$ contains, as a subring, the cohomology of $\mathcal X$.
\paragraph{}
Using the orbifold cohomology, we shall calculate the superconformal index, and
compare it to the localisation formulae. The quiver gauge theory we look at can
be found in figure \ref{C2Znquiver}. The unresolved
hyperK\"ahler cone is $\bC^2/\bZ_n$, which has an orbifold singularity at the
origin.

Chen Ruan cohomology involves taking the cohomology of the smooth part of the
manifold, as well as the addition of twisted sectors, which live at the
orbifold singularities. In the case of $\bC^2/\bZ_n$, there are $n-1$ twisted
sectors, corresponding to all non-identity elements\footnote{The twisted
sectors correspond to conjugacy classes in general, but this group is Abelian,
so they correspond to elements.} of $\bZ_n$. Each twisted sector is the point
set, $\{*\}$. So we have
\be
H^{p,q}_{\text{orb}}(\bC^2/\bZ_n)=H^{p,q}(\bC^2/\bZ_n)\oplus
H^{p-1,q-1}(\{*\})\,.
\ee
The ordinary cohomology bigraded-ring, $H^{p,q}(\bC^2/\bZ_n)$, is simply given
by taking $\bZ_n$-invariant holomorphic forms. $H^{0,0}(\{*\})=1$ and
$H^{p,q}(\{*\})=0$ for $(p,q)\neq(0,0)$.
\paragraph{}
Hence, we have that the superconformal index defined by the Chen Ruan
cohomology is
\be\label{tsu21}
\mathcal Z_{\text{orb}}(\bC^2/\bZ_n)=\frac1n\sum_{i=1}^{n}\frac\tau y\frac{1-y
(-z^2)^{i/n}}{1-\tau (-z^2)^{i/n}}\frac{1-y
/(-z^2)^{i/n}}{1-\tau
/(-z^2)^{i/n}}+n-1\,.
\ee

One can then test this against localisation formulae. Using the analysis of
section \ref{linearquiverchapter}, we know there are $n$ fixed points. Their
contribution to the index is
\be\label{tsu22}
\mathcal
Z(\bC^2/\bZ_n)=\sum_{a=1}^n\text{PE}\left[z^2\tau^{k-2a}\left(1-\frac\tau
y\right)(1-\tau y)\right]\,.
\ee
One can check that equations (\ref{tsu21}) and (\ref{tsu22}) are indeed the
same.
\paragraph{}
If we set $\tau=y$ to count the 1/2-BPS states, note that there is a
contribution of $n-1$ from the twisted sectors.

\subsection{$A$-type
quivers}\label{linearquiverchapter}\label{SCindexLinearADHMSection}

In this section, we compute the superconformal index for the case when
$\mathcal X$ is a Nakajima quiver variety of either $A$-type or $\hat A$-type.
We do this by using the known answer for instanton moduli space, an observation
by Nakajima in \cite{Nakajima:11}, that $A$-type and $\hat A$-type Nakajima
quiver varieties are $\bC^\times$-fixed point submanifolds of instanton moduli
space.

We explain the construction in \cite{Nakajima:11}. We take $\mathfrak
M_{\zeta_\bR,0}$ to be the resolved moduli space of $k$ $SU(N)$ instantons on
$\bC^2$ of section \ref{ADHMExample}. The construction takes a certain
$\bC^{\times}$-subgroup of $T:=\bC^\times\times G_H$, and restricts
$\mathfrak M_{\zeta_\bR,0}$ to the fixed point submanifold. This submanifold is
a disjoint union of linear quivers.
\paragraph{}
We want the fixed points of the $\bC^\times\ni t_1$ action on the set
$\mu_\bC^{-1}(0)$ given by
\be
(X,\tilde X,Q,\tilde Q)\mapsto (t_1X,t_1^{-1}\tilde
X,Q\rho_W(t_1)^{-1},\rho_W(t_1)\tilde Q)\,.
\ee
This corresponds to a choice of homomorphism $\rho_V:\mathbb C^\times\to
GL(\mathbb C^k)$, such that
\be\label{fixed}
(t_1X,t_1^{-1}\tilde X,Q\rho_W(t_1)^{-1},\rho_W(t_1)\tilde
Q)=(\rho_V(t_1)^{-1}X\rho_V(t_1),\rho_V(t_1)^{-1}\tilde
X\rho_V(t_1),\rho_V(t_1)^{-1}Q,\tilde Q\rho_V(t_1))\,.
\ee
$\rho_V$ is a homomorphism, because the action of $GL(\mathbb C^k)$ is free on
the space of $(X,\tilde X,Q,\tilde Q)$ that obey $\mu_\bR=\zeta_\bR$ for
$\zeta_\bR$ generic. A choice of conjugacy class of $\rho_V$ and $\rho_W$
defines a particular linear quiver. The conjugacy class of $\rho_V$ is
determined by an $\vec n\in \mathbb
Z^k/S_k$, such that
\be
t_1\mapsto\begin{pmatrix}t_1^{n_1}&&&\\&t_1^{n_2}&&\\&&\ddots&\\&&&t_1^{n_k}\end{pmatrix}\,.
\ee
Similarly, $\rho_W$'s conjugacy class is determined by $\vec m\in\mathbb
Z^N/S_N$.
\paragraph{}
We order these integers from smallest to largest. Let $p:=\min(m_1,n_1)$ and
$q:=\max(n_k,m_N)$. Define $n:=q-p+1$. For $a=1,\dots,n$, we define the spaces
\baa
V_a&=\text{Eigenspace of }\mathbb C^k\text{ with eigenvalue }t_1^{q+1-a}\,,\\
W_a&=\text{Eigenspace of }\mathbb C^N\text{ with eigenvalue }t_1^{q+1-a}\,.
\eaa
Note that unless $m_1\geqslant n_1$ and $m_N\leqslant n_k$, the fixed point set
will be empty, so we may as well take $n=n_k-n_1+1$.
\paragraph{}
The $\bC^\times$-fixed points respect the eigenspace structure of $\rho_V$ and
$\rho_W$: we see from equation (\ref{fixed}) that for $v\in V_i$,
\baa
&t_1Xv=\rho_V(t_1)^{-1}X\rho_V(t_1)v=t_1^{q+1-a}\rho_V(t_1)^{-1}Xv\,,\\
&\implies\rho_V(t_1)Xv=t_1^{q-a}Xv\,.
\eaa
This implies that $X:V_a\to V_{a+1}$. Similarly, $t_1^{-1}\tilde
X=\rho_V(t_1)^{-1}\tilde X\rho_V(t_1)$, $Q\rho_W(t_1)^{-1}=\rho_V(t_1)^{-1}Q$
and $\rho_W(t_1)\tilde Q=\tilde Q\rho_V(t_1)$ means
\baa
X&:V_a\to V_{a+1}\,,\\
\tilde X&:V_a\to V_{a-1}\,,\\
Q&:W_a\to V_a\,,\\
\tilde Q&:V_a\to W_a\,.
\eaa
So we exactly have the $A_n$ linear quiver. We define
\be
k_a:=\dim V_a\,,\quad N_a:=\dim W_a\,.
\ee

Calling the Nakajima quiver variety associated to the linear
quiver $M(\rho_V,\rho_W)$ -we have found that
\be
\coprod_{\rho_V}M(\rho_V,\rho_W)=\{\bC^\times\text{ fixed points of }\mathfrak
M_{\zeta_\bR,0}\}\,.
\ee

\label{FixedPointsATypeSection}

In the evaluation of the superconformal index of the linear quiver, the
analysis of section \ref{Localisation} means that one need only consider the
fixed points of the action of $T$. As discussed in section \ref{ADHMExample},
on instanton moduli space the fixed points correspond to $N$-coloured Young
tableaux of total size $k$. For a particular choice of $\rho_W$, each fixed
point will lie inside an individual linear quiver, corresponding to some
$\rho_V$. We explain here how to work out which
$\rho_V$, and hence which linear quiver, the fixed point is an element of. Note
that since the fixed points are invariant under the whole of
$(\bC^\times)^{N+2}$, they are invariant under the particular $\bC^\times$ we
used to restrict to the linear quivers, and hence must lie in some linear
quiver.
\paragraph{}
The Higgs branch of a linear quiver is non-empty if and only if it has a fixed
point under $T$. The only if is trivial, as the fixed point is
an element of the Higgs branch, while the other way is true because it must be
closed under the action of $(\bC^\times)^{N+2}$, lie within instanton moduli
space, and every point on instanton moduli space flows under the action of
$(\bC^\times)^{N+2}$ to a fixed point, \cite{Nakajima:03}.

\paragraph{}
 The fixed points are the maps $X,\tilde X, Q,\tilde Q$ such that
$\mu_\bR=\zeta_\bR$ and
$\mu_{\mathbb C}=0$, and
\baa
( \phi_l- \phi_m+\epsilon_1)X_{lm}&=0\,,\\
( \phi_l-\phi_m+\epsilon_2)\tilde X_{lm}&=0\,,\\
( \phi_l-\frac{\epsilon_1+\epsilon_2}{2} -a_i)Q_{li}&=0\,,\\
( \phi_l+\frac{\epsilon_1+\epsilon_2}{2} -a_i)\tilde Q_{il}&=0\,,
\eaa
where $l,m=1,\dots,k$ and $i=1,\dots,N$, and the $(\phi_l)_l$ are diagonalised
gauge transformations. The coloured Young tableaux give us a way of reindexing
the numbers $l=1,\dots,k$ as $(i,(\alpha,\beta))$ for $(\alpha,\beta)\in Y_i$
and $i=1,\dots,N$. 2Exactly $k$ components of the $2kN+2k^2$ components of
$(X,\tilde X,Q,\tilde
Q)$ are non-zero. They are
\be
X_{i(\alpha,\beta),i(\alpha+1,\beta)}, \tilde
X_{i(\alpha,\beta),i(\alpha,\beta+1)}, \tilde Q_{i(1,1),i}\neq0\,.
\ee

\begin{figure}[h]
\centering
\begin{tikzpicture}[]
\node[minimum size=0.6cm,draw, circle](lin1) at (15,-1.6){$1$};
\node[minimum size=0.6cm,draw, circle](lin2) at (16.5,-1.6){$2$};
\node at (18,-1.6) {$\ldots$};
\node[minimum size=0.6cm,draw, circle](lin3) at (19.5,-1.6){$l$};
\node at (21,-1.6) {$\ldots$};
\node[minimum size=0.6cm,draw, circle](lin4) at (22.5,-1.6){$2$};
\node[minimum size=0.6cm,draw, circle](lin5) at (24,-1.6){$1$};
\node[minimum size=0.7cm,draw](Fin1) at (19.5,0){$N$};
\draw[-](lin1.east)--(lin2.west);
\draw[-](lin4.east)--(lin5.west);
\draw[-](lin2.east)--(17.5,-1.6);
\draw[-](lin3.west)--(18.5,-1.6);
\draw[-](lin3.north)--(Fin1.south);
\draw[-](lin3.east)--(20.5,-1.6);
\draw[-](lin4.west)--(21.5,-1.6);
\end{tikzpicture}
\caption{An example of a linear quiver. It is a connected component of the
resolved moduli space of $k=l^2$ $SU(N)$ instantons on
$\bC^2$.}\label{linearfig}
\end{figure}
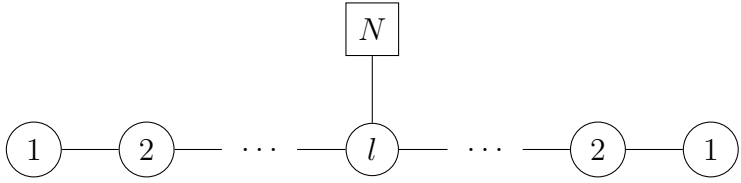

Suppose $e_{ais}$ is a basis for $\mathbb C^k$, for $a=1,\dots,n\,,$
$i=1,\dots,N_a$ and $s\in Y_{ai}$, and $f_{ai}$ a basis for $\mathbb C^{N_a}$
for $a=1,\dots,n$, $i=1,\dots, N_a$. Then we have that $e_{ai(1,1)}\in V_a$,
because $\tilde Q_{ai(1,1)\,ai}f_{ai}\propto e_{ai(1,1)}$ and $\tilde
Q_{ai(1,1)\,ai}\neq0$. Now we see that if $(2,1)\in Y_{ai}$, then
$Xe_{ai(2,1)}\propto e_{ai(1,1)}$, and so $e_{ai(2,1)}\in V_{a-1}$. Through
this, we see that
\be
e_{ai(\alpha,\beta)}\in V_{a-\alpha+\beta}\,.
\ee
This fully determines the value of the $k_a$'s. Note that there can be values
of $a$ where $N_a=0$ and $k_a\neq0$.
\paragraph{}
A special class of linear quivers are known as $T_\sigma^\rho\big(SU(M)\big)$
quivers.  $\sigma$ and $\rho$ are partitions of $M$ determined by $\rho_W$ and
$\rho_V$. The $\rho$ and $\sigma$ are defined by $\rho_i^\vee-\rho^\vee_i=N_i$
and $k_i=\rho^\vee_1+\dots+\rho^\vee_i-\sigma_1-\dots-\sigma_i$. The quiver
will not be a $T^\rho_\sigma$ if the gauge group ranks do not define a
partition through the equation for $k_i$ before. However, if this is the case,
the quiver is Seiberg dual to a $T^\rho_\sigma$, see appendix
\ref{WallCrossingAppendix}. The
number of fixed points on the resolved space for a
$T_\sigma^\rho\big(SU(M)\big)$ is
\be
\#\text{ of fixed
points}=\sum_{\rho\leqslant\nu\leqslant\sigma^\vee}K_{\nu\rho}K_{\nu^\vee\sigma}\,,
\ee
where $K_{\alpha\beta}$ are the Kostka numbers. This is due to the expression
for the Poincar\'e polynomial of $A$-type Nakajima quiver varieties in
\cite{Proudfoot:2014}, and the fact that each fixed point contributes a
generator to the homology.
\paragraph{}
We look at an example, to show how the Young tableaux are chosen. Take the
linear quiver in figure \ref{linearfig}

The coloured Young tableau that we restrict to are exactly the ones such that
\baa
&\exists\text{ exactly }l\,,s\in Y_i\text{ s.t. }s=(b,b)\,\text{for some
}b\in\mathbb N\,,\\
&\exists\text{ exactly }l-1\,,s\in Y_i\text{ s.t. }s=(b-1,b)\,\text{for some
}b\in\mathbb N\,,\\
&\exists\text{ exactly }l-1\,,s\in Y_i\text{ s.t. }s=(b,b-1)\,\text{for some
}b\in\mathbb N\,,\\
&\exists\text{ exactly }l-2\,,s\in Y_i\text{ s.t. }s=(b-2,b)\,\text{for some
}b\in\mathbb N\,,\\
&\text{etc}\,.
\eaa
So if $N=1$, the only pole is given by a single square Young tableaux of height
and width $l$.

\paragraph{}

\label{LimitoftheIndex}

We write the superconformal index for the linear quiver defined by the
conjugacy classes of $\rho_V$ and $\rho_W$. We restrict to the fixed points
corresponding to the linear quiver, scale the fugacities according to the
$\bC^\times$-action and take the limit $x\to0$.
This gives
\baa\label{Nekpartfnlin}
\mathcal Z_{\rho_V,\rho_W}
&=\!\!\sum_{\substack{\{Y_{a,i}\}\\\rho_V,\,\rho_W}}\!\!\prod_{a,b=1}^n\prod_{i=1}^{N_a}\prod_{j=1}^{N_b}\!\!\!\!\!\!\!\!\!\prod_{\substack{s\in
Y_{a,i}\\f_{(a,i)(b,j)}(s)=a-b}}\!\!\!\!\!\!\!\!\!\!\!\text{PE}\left[\frac{z_{a,i}}{z_{b,j}}\tau^{g_{(a,i)(b,j)}(s)-1}(1-\tau/y)(1-\tau
y)\right]\,.
\eaa

In this expression, the $\substack{\{Y_{a,i}\}\\\rho_V,\rho_W}$ means
restricting the sum to all fixed points corresponding to the linear quiver
$M(\rho_V,\rho_W)$. Note that unlike the instanton moduli space's
superconformal index, a generic box from a coloured Young tableaux, associated
to a fixed point within the manifold, need not contribute an individual term to
the index. Indeed, if this were so then the highest power of $y$ in the index
would be $kN$, which is strictly greater than the quaternionic dimension of
$M(\rho_V,\rho_W)$.
\paragraph{}
Since the manifold is connected, and the contribution at each fixed point
corresponds to the tangent space at that point, we would expect that the
highest power of $y$ at each point would be the quaternionic dimension of the
manifold $\dim_{\bH}M(\rho_V,\rho_w)=\sum_a(k_ak_{a+1}+k_aN_a-k_a^2)$. This is
a non-trivial combinatorial condition on the coloured Young tableaux that
appears to be true.
\paragraph{}

From this we can conclude that
\be\label{ADHMlimitgivesLinears}
\lim_{\rho_W,\,x\to0}\mathcal Z_{k,N}=\sum_{\rho_V}\mathcal
Z_{\rho_V,\rho_W}\,.
\ee
This sum has multiplicity one for each $\rho_V$, but we might find that
$\mathcal Z_{\rho_V,\rho_W}=0$, and we may also have two equivalent linear
quivers for different $\rho_V$'s, for example (1)-(1)-[1] and [1]-(1)-(1).
Furthermore, we may not have a connected quiver for a specific $\rho_V$ and
$\rho_W$.

\subsubsection{An example: cotangent bundles of flag
varieties}\label{needanamefortheselinearquivers}
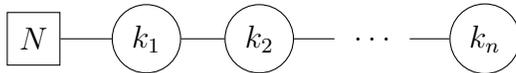
\begin{figure}
\centering
\begin{tikzpicture}[]
\node[minimum size=0.6cm,draw, circle](lin3) at (19.5,-1.6){$k_1$};
\node at (22.5,-1.6) {$\ldots$};
\node[minimum size=0.6cm,draw, circle](lin4) at (21,-1.6){$k_{2}$};
\node[minimum size=0.6cm,draw, circle](lin5) at (24,-1.6){$k_n$};
\node[minimum size=0.7cm,draw](flav) at (18,-1.6){$N$};
\draw[-](23,-1.6)--(lin5.west);
\draw[-](lin3.west)--(flav.east);
\draw[-](lin3.east)--(lin4.west);
\draw[-](lin4.east)--(22,-1.6);
\end{tikzpicture}
\caption{A quiver whose resolved space is a cotangent bundle to a flag
variety.}\label{anotherlinearfig}
\end{figure}

We look at a special class of linear quivers, the ones in figure
\ref{anotherlinearfig}.

The quiver ranks obey  $N\geqslant k_1\geqslant k_2\geqslant\dots\geqslant
k_n>0$, otherwise the Higgs branch is empty. The unresolved space is a
nilpotent orbit, while the resolved space is the cotangent bundle to the flag
variety
$\{\bC^{k_n}\hookrightarrow\bC^{k_{n-1}}\hookrightarrow\dots\hookrightarrow\bC^N\}$.
The calculation of the Hilbert
series of this quiver was done via Lefschetz fixed point theorem directly in
\cite{Hanany:14}, we find that our analysis exactly reproduces their results
for a choice of $k$ and $N$ such that it is a $T_\rho(SU(N))$ theory.

\paragraph{}
 Define the composition of $N$
\be
l_1=k_n\,,\quad l_2=k_{n-1}-k_n\,,\quad l_3=k_{n-2}-k_{n-1}\,,\dots\,,
l_n=k_1-k_2\,,\quad l_{n+1}=N-k_1\,.
\ee

Since there is only one flavour node, and it is on the far left node, the fixed
points are coloured Young tableaux of length 1. They are given by $l_1$ lots of
$(n)$, $l_2$ lots of $(n-1)$,\dots, $l_n$ lots of $(1)$ and $l_{n+1}$
lots of $\varnothing$. One then needs to sum over the Weyl group $S_N$ modulo
the Weyl group of the Levi subgroup,
\be
\tilde W:=\prod_{a=1}^{n+1}S_{l_a}\,.
\ee
This is precisely the same parameterisation of the fixed points found in
\cite{Hanany:14}. The quaternionic dimension of the
manifold is
\be
d_H:=\sum_{a> b}l_al_b\,.
\ee

Suppose $Y_i=(a)$ and $Y_j=(b)$, then we have that $f_{ij}(s)$ is zero if and
only if $b\leqslant a-1$ and $s=(1,a)$ (the last box). For this box we have
that $g_{ij}(s)=-1$. Define the function on indices
$h:\{1,\dots,N\}\to\{0,1,\dots,n\}$ via
\be\label{defineh}
i=l_1+l_2+\dots+l_{h(i)}+j\,,\text{ for }j=1,\dots,l_{h(i)+1}\,.
\ee

The superconformal index is
\baa\label{fish-tail}
\mathcal Z&=\sum_{w\in S_N/\tilde W}\prod_{h(i)>h(j)}w\left(\frac
\tau{y}\frac{\left(1-\frac{z_i}{z_j}\frac
y\tau\right)\left(1-\frac{z_j}{z_i}\tau
y\right)}{\left(1-\frac{z_i}{z_j}\right)\left(1-\frac{z_j}{z_i}\tau^{2}\right)}\right)\,.
\eaa

The orbit under Seiberg duality of the quiver in figure \ref{anotherlinearfig}
is a family of the same structure, but with varying gauge ranks. There is an
element of this orbit where $(l)$, the corresponding composition of $N$,  is in
fact a decreasing sequence, and so defines a partition
$\rho=(l_{n+1},l_n,\dots,l_1)$. This quiver is the $T_\rho(SU(N))$-quiver.

Note that the resolved space is the cotangent bundle to a flag variety, with
$\tau$ the grading for the $\bC^\times$-action rotating the cotangent fibre.
This means that if we set $\tau\to0$ then we restrict to the flag variety
itself. The flag variety is projective and is a deformation retract of its
cotangent bundle. This means that the $\tau\to0$ limit of the superconformal
index is the Poincar\'e polynomial of the cotangent bundle to the flag variety,
giving
\baa
P_{T_\rho(SU(N))}(\tilde y)&=\sum_{\sigma\in S_N/\tilde
W}\sigma\left(\prod_{h(i)>h(j)}\frac{1-\tilde
y\frac{z_i}{z_j}}{1-\frac{z_i}{z_j}}\right)\,.
\eaa

Using our own limit for the Poincar\'e polynomial we obtain
\baa
P_{T_\rho(SU(N))}(\tilde y)&=\sum_{\sigma\in S_N/\tilde W}\tilde
y^{\ell(\sigma)}\,,
\eaa
where $\ell(\sigma)$ is the length of the shortest element of the coset
$\sigma\tilde
W$.

Finally, it is also known classically that the Poincar\'e polynomial of the
flag variety is
\baa
P_{T_\rho(SU(N))}(\tilde
y)&=\sum_{\nu\geqslant\rho}K_{\nu^\vee\,(1^N)}K_{\nu\rho}(\tilde y)\\
&=\frac{\prod_{i=1}^N(1-\tilde
y^i)}{\prod_{j=1}^{\ell(\rho)}\prod_{i=1}^{\rho_j}(1-\tilde y^i)}\,.
\eaa

It is proven in \cite{Mythesis} that any
linear quiver's superconformal index can be reached by taking a certain limit
of the flavour fugacities of a linear quiver whose resolved space is the
cotangent bundle to a flag variety. This means we can write the superconformal
index of any linear quiver as a single Weyl group sum.

\subsection{$\hat A$-type quivers}\label{affine A quivers}
The construction of $A_n$-type quivers can be easily adapted to give us the
fixed points of generic $\hat A_n$-type quivers. The associated variety to a
$\hat A_n$ quiver is the moduli space of instantons on $\bC^2/\bZ_n$,
\cite{Kronheimer:1990}.
\paragraph{}
We restrict $t_1$ in equation (\ref{fixed}) to lie in a finite cyclic group,
$t_1\in\bZ_n\subset\bC^\times$. If we do this, then we have the same argument
as for the linear quiver. $\bC^k$ is split into $n$ pieces,
$V_0,\dots,V_{n-1}$, with $V_a$ having weight $t_1^a$, and similarly $\bC^N$
splits into $n$ pieces $W_0,\dots,W_{n-1}$. As before we define
\be
k_a:=\dim\,V_a\,,\quad N_a:=\dim\,W_a\,.
\ee
The difference now is the periodicity, namely
\baa
X&:V_{n-1}\to V_0\,,\\
\tilde X&:V_0\to V_{n-1}\,.
\eaa

This periodicity is important for identifying which $\hat A_n$-type quiver a
particular fixed point lies in given the choice of $\rho_W$.
\paragraph{}
The superconformal index for $\hat A_n$ is
\baa\label{Nekpartfnaff}
&\mathcal Z_{\rho_V,\rho_W}(\hat A_n)\\
=&\sum_{\substack{
\{Y_{a,i}\}\\\rho_V,\,\rho_W}}\prod_{a,b=1}^n\prod_{i=1}^{N_a}\prod_{j=1}^{N_b}\!\!\!\!\!\!\!\!\!\!\prod_{\substack{s\in
Y_{a,i}\\f_{(a,i)(b,j)}(s)\equiv
a-b\,(\text{mod
}n)}}\!\!\!\!\!\!\!\!\!\!\!\!\text{PE}\left[\frac{z_{ai}}{z_{bj}}\tau^{g_{(a,i),(b,j)}(s)-1}(1-\tau/y)(1-\tau
y)\right]\,.
\eaa
In this expression, the $\substack{\{Y_{a,i}\}\\\rho_V,\rho_W}$ means
restricting the sum to all fixed points corresponding to the affine quiver
fixed by $\rho_V$ and $\rho_W$.

\section*{Acknowledgements}   
The authors would like to thank Sam Crew, David Tong, Joe Waldron, Tom\'a\u{s} Zemen and Filip Zivanovic  for helpful discussions. This paper has been partially supported by STFC grant ST/L000385/1.

\appendix
\section{Some superalgebra details}\label{AppendixOSp}

\subsection{Flat space quantum mechanics}
We consider the problem of free quantum mechanics on $\bC^{2n}$, with complex
coordinates $(q_i,\tilde q_i)_{i=1}^n$. Consider the action of
$-\bM-2\bJ_3+\bN$ on
the forms
\be
\alpha=\prod_{i=1}^nq_i^{a_i}\bar q_i^{\bar a_i}\tilde q_i^{b_i}\bar{\tilde
q}_i^{\bar b_i}dq_i^{\delta_i}\wedge d\bar q_i^{\bar\delta_i}\wedge d\tilde
q_i^{\epsilon_i}\wedge d\bar{\tilde q}_i^{\bar\epsilon_i}\,,
\ee
with $a,\bar a,b,\bar b\in\bZ_{\geqslant0}^n$ and
$\delta,\bar\delta,\epsilon,\bar\epsilon\in\{0,1\}^n$, then we have
that\footnote{Explicit expressions for the $\mathfrak{osp}(4^*|4)$ generators
on flat space can be found in appendix E.5 of \cite{Singletonthesis}.}
\baa
2\bJ_3\alpha&=\sum_{i=1}^n(a_i+b_i-\bar a_i-\bar b_i)\alpha\,,\\
\bM\alpha&=\sum_{i=1}^n(\bar\delta_i+\bar\epsilon_i-n)\alpha\,,\\
\bN\alpha&=\sum_{i=1}^n(\delta_i+\epsilon_i-n)\alpha\,.
\eaa
So, $-\bM-2\bJ_3+\bN$ acts as the $\bC^\times$-scaling\footnote{On flat space,
this
is the Lie derivative with respect to the Hamiltonian vector field
$\sum_{i=1}^n\left(q_i\frac{\partial}{\partial q_i}+\tilde
q_i\frac{\partial}{\partial \tilde q_i}-\bar q_i\frac{\partial}{\partial \bar
q_i}-\bar{\tilde q}_i\frac{\partial}{\partial \bar{\tilde q}_i}\right)$.},
which we call $R$:
\be
R\alpha=(-\bM-2\bJ_3+\bN)\alpha=\sum_{i=1}^n(a_i+b_i-\bar a_i-\bar
b_i+\delta_i+\epsilon_i-\bar\delta_i-\bar\epsilon_i)\alpha\,.
\ee

\subsection{$\mathfrak{su}(1|2)$ character}\label{su21characterexpansion}
Suppose that $p(\tau,y)\in\bZ((Z))[[\tau]][y,1/y]$ is such that
\be\label{propertya4}
p(\tau,y,Z)=p(\tau,1/y,Z)\,,
\ee
a formal Laurent series in $Z$, a formal power series in $\tau$ and a finite
Laurent expansion in $y$, writing it as
\be
p(\tau,y,Z)=\sum_{a=0}^\infty\sum_{b=-m}^m\alpha_{a,b}(Z)\tau^ay^b=\sum_{a=0}^\infty\sum_{b=-m}^m\alpha_{a,b}(Z)\tau^{a+b}\left(\frac
y\tau\right)^b\,,
\ee
we further have that
\be\label{propertya6}
\alpha_{a,b}=0\text{ for }a<|b|\,,
\ee
that
\be\label{propertya7}
p(\tau,\tau,Z)\in\bZ_{\geqslant0}\,,\text{ so that
}\frac{d}{d\tau}p(\tau,\tau,Z)=\frac{d}{dz_i}p(\tau,\tau,Z)=0\,,
\ee
and
\be\label{propertya8}
\lim_{\substack{\tau\to0\\y/\tau\text{
finite}}}p(\tau,y,Z)=\sum_{a=0}^m\alpha_{a,-a}(Z)\left(\frac\tau
y\right)^a\in\bZ_{\geqslant0}\left[\frac \tau y\right]\,.
\ee
We have that in the text equation (\ref{propertya4}) corresponds to (\ref{3propertya4}), equation (\ref{propertya6}) corresponds to (\ref{3propertya6}), equation (\ref{propertya7}) corresponds to (\ref{3propertya7}) and equation (\ref{propertya8}) corresponds to (\ref{3propertya8})

\paragraph{}

We show in this subsection that if $p$ obeys all these properties, then it can
be written as
\be
p(\tau,y)=\sum_{a=0}^mN_{a,a}\mathcal I_{a,a}+\sum_{a>b}\tilde N_{a,b}\mathcal
I_{a,b}\,,
\ee
with $N_{a,a}\in\bZ_{\geqslant0}$ and $\tilde N_{a,b}\in\bZ[Z,Z^{-1}]$.

We prove this via induction on $m$. If $m=0$, then since
$p(\tau,\tau,Z)=p(\tau,y,Z)=\alpha_{0,0}\in\bZ_{\geqslant0}$ and $\mathcal
I_{0,0}=1$, we have
\be
p(\tau,y)=\alpha_{0,0}\mathcal I_{0,0}\,.
\ee

Now suppose it is true up to the highest power of $y$ in $p$ being $m-1$. We
take
\be
p(\tau,y,Z)=q(\tau,Z)(y^m+1/y^m)+\sum_{a=0}^\infty\sum_{b=1-m}^{m-1}\alpha_{a,b}(Z)\tau^ay^b\,,
\ee
for some $q(\tau,Z)\in\tau^m\bZ((Z))[[\tau]]$, $q(\tau,Z)=\sum_{a=m}^\infty
q_v(Z)\tau^v$. Equation (\ref{propertya8}) necessarily means that
$q_m(Z)\in\bZ_{\geqslant0}$. We see that we can write
\be
p(\tau,y,Z)=q_m\mathcal I_{m,m}-\sum_{a=m+1}^nq_a(Z)\mathcal
I_{a-1,m-1}(\tau,y)+\tilde p(\tau,y,Z)\,,
\ee
where the highest power of $y$ in $\tilde p(\tau,y)$ is $m-1$ and it clearly
obeys all the necessary properties (\ref{propertya4}), (\ref{propertya6}),
(\ref{propertya7}) and (\ref{propertya8}).
\section{Wall crossing}\label{WallCrossingAppendix}
We conjecture that the superconformal index does not depend on the choice of
projective symplectic resolution. We discuss this further in this appendix for
certain Nakajima quiver varieties, providing some of the evidence for this
conjecture.

\paragraph{}
Define $\Gamma=(V,\Omega)$ to be a quiver of $ADE$-type. If $\omega_i$ are the
fundamental weights  and $\alpha_i$ are the simple roots of $\Gamma$. Then for
a given choice of gauge ranks $k\in\bZ_{>0}^V$, flavour ranks
$N\in\bZ_{\geqslant0}^V$, choice
of FI parameter $\zeta\equiv\zeta_\bR\in\bR^V$ and background baryonic charge
$B\in\bZ^V$, we define
\baa
\lambda&:=\sum_{i\in V}N_i\omega_i\,,\quad\alpha:=\sum_{i\in V}k_i\alpha_i\,,\\
\zeta&:=\sum_{i\in V}\zeta_i\omega_i\,.
\eaa

There is an action of the Weyl group of the
quiver, $W\equiv W_\Gamma$. For $w\in W$:
\baa
\zeta&\mapsto w(\zeta),\quad B\mapsto w(B)\,,\\
\alpha&\mapsto w*\alpha:= \lambda-w(\lambda-\alpha)\,.
\eaa

From the work  \cite{Nakajima:2003reflection}, we know that
\be
\mathfrak M(\lambda,\alpha,\zeta)\cong \mathfrak
M(\lambda,w*\alpha,w(\zeta))\,,
\ee
where the isomorphism is a hyperK\"ahler isometry, an isometry preserving all
three complex structures. This transformation is known in the mathematical
literature as reflection functors, while in the physical literature it is known
as three dimensional Seiberg duality.
\paragraph{}

The Hilbert series is known to be independent of the choice of $\zeta$,
\cite{Ginzburg:2009lectures}. However, in general the fixed point cotangent
space $T$-module structure does depend on the resolution.

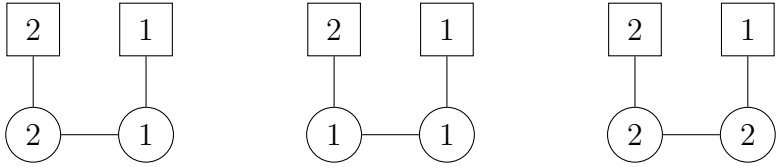
\begin{figure}[h]
\centering
\begin{tikzpicture}[]
\node[minimum size=0.6cm,draw, circle](lin1) at (0,0){$2$};
\node[minimum size=0.6cm,draw, circle](lin2) at (1.5,0){$1$};
\node[minimum size=0.7cm,draw](Fin1) at (0,1.4){$2$};
\node[minimum size=0.7cm,draw](Fin2) at (1.5,1.4){$1$};
\draw[-](lin1.east)--(lin2.west);
\draw[-](lin1.north)--(Fin1.south);
\draw[-](lin2.north)--(Fin2.south);
\node[minimum size=0.6cm,draw, circle](lin11) at (4,0){$1$};
\node[minimum size=0.6cm,draw, circle](lin12) at (5.5,0){$1$};
\node[minimum size=0.7cm,draw](Fin11) at (4,1.4){$2$};
\node[minimum size=0.7cm,draw](Fin12) at (5.5,1.4){$1$};
\draw[-](lin11.east)--(lin12.west);
\draw[-](lin11.north)--(Fin11.south);
\draw[-](lin12.north)--(Fin12.south);
\node[minimum size=0.6cm,draw, circle](lin21) at (8,0){$2$};
\node[minimum size=0.6cm,draw, circle](lin22) at (9.5,0){$2$};
\node[minimum size=0.7cm,draw](Fin21) at (8,1.4){$2$};
\node[minimum size=0.7cm,draw](Fin22) at (9.5,1.4){$1$};
\draw[-](lin21.east)--(lin22.west);
\draw[-](lin21.north)--(Fin21.south);
\draw[-](lin22.north)--(Fin22.south);
\end{tikzpicture}
\caption{We investigate the quiver on the left. For this quiver
$\lambda=2\omega_1+\omega_2$ and $\lambda-\alpha=\omega_3+\omega_2-\omega_1$.
The quiver in the middle is the image under the Weyl group transformation
$(12)$, and has $\lambda-(12)*\alpha=\omega_3+\omega_1$. The quiver on the
right is the image under $(23)$ and has
$\lambda-(23)*\alpha=2\omega_3-\omega_2$. The middle quiver is the
$T^{(3,1)}_{(2,1,1)}(SU(4))$ quiver.}\label{lengthtwoquiver}
\end{figure}

For example, the polynomial
$p_x(\tau,Z)\in\bZ_{\geqslant0}[\tau^{\pm1},Z^{\pm1}]$ from
equation (\ref{FixedPointPolynomial}) for the quiver on the left in figure
\ref{lengthtwoquiver} has the following values for the five fixed points of the
two resolutions:
\baa
p_x\text{ for }\zeta^{(1)}=(1,1):
&\frac{z_{1,1}}{z_{1,2}}+\tau\frac{z_{1,1}}{z_{2,1}},\quad
\frac{z_{1,2}}{z_{1,1}}+\tau\frac{z_{1,2}}{z_{2,1}},\quad\tau^{-1}\frac{z_{1,1}}{z_{2,1}}+\tau^{-1}\frac{z_{1,2}}{z_{2,1}},\\
&\frac{z_{1,2}}{z_{1,1}}+\tau\frac{z_{2,1}}{z_{1,1}},\quad\frac{z_{1,1}}{z_{1,2}}+\tau\frac{z_{2,1}}{z_{1,2}}\,,\\
p_x\text{ for }\zeta^{(2)}=(2,-1):
&\tau\frac{z_{1,2}}{z_{2,1}}+\tau\frac{z_{1,1}}{z_{2,1}},\quad
\frac{z_{1,2}}{z_{1,1}}+\tau^{-1}\frac{z_{1,1}}{z_{2,1}},\quad\frac{z_{1,1}}{z_{1,2}}+\tau^{-1}\frac{z_{1,2}}{z_{2,1}},\\
&\frac{z_{1,2}}{z_{1,1}}+\tau\frac{z_{2,1}}{z_{1,1}},\quad\frac{z_{1,1}}{z_{1,2}}+\tau\frac{z_{2,1}}{z_{1,2}}\,.
\eaa

We see that the fixed point structure is different for these two different
choices of resolution. Nonetheless, both the Hilbert series and the
superconformal index are the same. We have tested this for multiple length two
quivers and some length three quivers and the same structure is persistent.
\paragraph{}
Note that $\zeta^{(1)}=(23)(\zeta^{(2)})$, while the subgroup of the Weyl group
of the quiver, $W_{A_2}=S_3$, for which $\mu$ is invariant, namely the group
$\{1,(13)\}\cong\bZ_2$, the fixed point structure is invariant. This holds more
generally, as the resolutions are Seiberg dual to each other (equivalently one
is given by acting with a reflection functor on the other). This further means
that for the quiver on the right in figure \ref{lengthtwoquiver}, the fixed
point structure for the left quiver with choice of resolution
$\zeta_{\text{left}}$ is equal to the fixed point structure for the right
quiver with resolution $\zeta_{\text1{right}}:=(23)(\zeta_{\text{left}})$.

\bibliographystyle{plain}

\cleardoublepage
\bibliography{references}

\begin{thebibliography}{10}

\bibitem{Aharony:1997matrix}
Ofer Aharony, M~Berkooz, S~Kachru, N~Seiberg, and Eva Silverstein.
\newblock Matrix description of interacting theories in six dimensions.
\newblock {\em arXiv preprint hep-th/9707079}, 1997.

\bibitem{Aharony:1997light}
Ofer Aharony, Micha Berkooz, and Nathan Seiberg.
\newblock Light-cone description of (2, 0) superconformal theories in six
  dimensions.
\newblock {\em arXiv preprint hep-th/9712117}, 1997.

\bibitem{Alvarez:1981geometrical}
Luis {\'A}lvarez-Gaum{\'e} and Daniel~Z Freedman.
\newblock Geometrical structure and ultraviolet finiteness in the
  supersymmetric $\sigma$-model.
\newblock {\em Communications in Mathematical Physics}, 80(3):443--451, 1981.

\bibitem{Atiyah:1994construction}
Michael~F Atiyah, Nigel~J Hitchin, Vladimir~Gershonovich Drinfeld, and Yu~I
  Manin.
\newblock Construction of instantons.
\newblock In {\em Instantons In Gauge Theories}, pages 133--135. World
  Scientific, 1994.

\bibitem{Atiyah:1968index}
Michael~F Atiyah and Graeme~B Segal.
\newblock The index of elliptic operators: {II}.
\newblock {\em Annals of Mathematics}, pages 531--545, 1968.

\bibitem{Mythesis}
Alec~E. Barns-Graham.
\newblock Much ado about nothing: {T}he superconformal index and {H}ilbert
  series of three dimensional {$\mathcal N=4$} vacua.
\newblock PhD thesis, 2018.

\bibitem{Chen:2004}
Weimin Chen and Yongbin Ruan.
\newblock A new cohomology theory of orbifold.
\newblock {\em Communications in Mathematical Physics}, 248(1):1--31, 2004.

\bibitem{Cremonesi:2014coulomb}
Stefano Cremonesi, Giulia Ferlito, Amihay Hanany, and Noppadol Mekareeya.
\newblock Coulomb branch and the moduli space of instantons.
\newblock {\em Journal of High Energy Physics}, 2014(12):103, 2014.

\bibitem{Hanany:14}
Stefano Cremonesi, Amihay Hanany, Noppadol Mekareeya, and Alb~erto Zaffaroni.
\newblock {{$T_\rho^\sigma (G)$} theories and their {H}ilbert series}.
\newblock {\em JHEP}, 01:150, 2015.

\bibitem{Hanany:14glue}
Stefano Cremonesi, Amihay Hanany, Noppadol Mekareeya, and Alberto Zaffaroni.
\newblock {Coulomb branch {H}ilbert series and {H}all-{L}ittlewood
  polynomials}.
\newblock {\em JHEP}, 09:178, 2014.

\bibitem{Cremonesi:2014sicilian}
Stefano Cremonesi, Amihay Hanany, Noppadol Mekareeya, and Alberto Zaffaroni.
\newblock Coulomb branch {H}ilbert series and three dimensional {S}icilian
  theories.
\newblock {\em Journal of High Energy Physics}, 2014(9):185, 2014.

\bibitem{Cremonesi:2014monopole}
Stefano Cremonesi, Amihay Hanany, and Alberto Zaffaroni.
\newblock Monopole operators and {H}ilbert series of {C}oulomb branches of 3d
  {$\mathcal N=4$} gauge theories.
\newblock {\em Journal of High Energy Physics}, 2014(1):5, 2014.

\bibitem{de:1976conformal}
Vittorio de~Alfaro, S~Fubini, and G~Furlan.
\newblock Conformal invariance in quantum mechanics.
\newblock {\em Il Nuovo Cimento A (1965-1970)}, 34(4):569--612, 1976.

\bibitem{de2001:hypermultiplets}
Bernard de~Wit, Martin Rocek, and Stefan Vandoren.
\newblock Hypermultiplets, hyperk{\"a}hler cones and quaternion-{K\"a}hler
  geometry.
\newblock {\em Journal of High Energy Physics}, 2001(02):039, 2001.

\bibitem{Dorey:2018}
Nick Dorey and Andrew Singleton.
\newblock An index for superconformal quantum mechanics.
\newblock To be published, 2018.

\bibitem{Ginzburg:2009lectures}
Victor Ginzburg.
\newblock Lectures on {N}akajima's quiver varieties.
\newblock {\em arXiv preprint arXiv:0905.0686}, 2009.

\bibitem{Griffiths:2014principles}
Phillip Griffiths and Joseph Harris.
\newblock {\em Principles of algebraic geometry}.
\newblock John Wiley \& Sons, 2014.

\bibitem{Hartshorne:2013algebraic}
Robin Hartshorne.
\newblock {\em Algebraic geometry}, volume~52.
\newblock Springer Science \& Business Media, 2013.

\bibitem{Hatayama:1998}
G.~Hatayama, A.~Kuniba, M.~Okado, T.~Takagi, and Y.~Yamada.
\newblock Remarks on fermionic formula.
\newblock In {\em Recent developments in quantum affine algebras and related
  topics ({R}aleigh, {NC}, 1998)}, volume 248 of {\em Contemp. Math.}, pages
  243--291. Amer. Math. Soc., Providence, RI, 1999.

\bibitem{Hausel:05}
Tam\'as Hausel.
\newblock Betti numbers of holomorphic symplectic quotients via arithmetic
  {F}ourier transform.
\newblock {\em Proc. Natl. Acad. Sci. USA}, 103(16):6120--6124, 2006.

\bibitem{Hausel:08}
Tam\'as Hausel.
\newblock Kac's conjecture from {N}akajima quiver varieties.
\newblock {\em Invent. Math.}, 181(1):21--37, 2010.

\bibitem{Jeffrey:1995localization}
Lisa~C Jeffrey and Frances~C Kirwan.
\newblock Localization for nonabelian group actions.
\newblock {\em Topology}, 34(2):291--327, 1995.

\bibitem{Kaledin:2003crepant}
D~Kaledin.
\newblock On crepant resolutions of symplectic quotient singularities.
\newblock {\em Selecta Mathematica, New Series}, 9(4):529--555, 2003.

\bibitem{Kim:11}
Hee-Cheol Kim, Seok Kim, Eunkyung Koh, Kimyeong Lee, and Sungjay Lee.
\newblock {On instantons as {K}aluza-{K}lein modes of {M5}-branes}.
\newblock {\em JHEP}, 12:031, 2011.

\bibitem{Kronheimer:1990}
Peter~B Kronheimer and Hiraku Nakajima.
\newblock Yang-{M}ills instantons on {ALE} gravitational instantons.
\newblock {\em Mathematische Annalen}, 288(1):263--307, 1990.

\bibitem{Michelson:2000geometry}
Jeremy Michelson and Andrew Strominger.
\newblock The geometry of (super) conformal quantum mechanics.
\newblock {\em Communications in Mathematical Physics}, 213(1):1--17, 2000.

\bibitem{Mozgovoy:2006}
Sergey Mozgovoy.
\newblock Fermionic forms and quiver varieties.
\newblock 2006.

\bibitem{Nakajima:1999lectures}
Hiraku Nakajima.
\newblock {\em Lectures on {H}ilbert schemes of points on surfaces}.
\newblock Number~18. American Mathematical Soc., 1999.

\bibitem{Nakajima:2001quiver}
Hiraku Nakajima.
\newblock Quiver varieties and finite dimensional representations of quantum
  affine algebras.
\newblock {\em Journal of the American Mathematical Society}, 14(1):145--238,
  2001.

\bibitem{Nakajima:2003reflection}
Hiraku Nakajima.
\newblock Reflection functors for quiver varieties and {W}eyl group actions.
\newblock {\em Mathematische Annalen}, 327(4):671--721, 2003.

\bibitem{Nakajima:11}
Hiraku Nakajima.
\newblock Handsaw quiver varieties and finite {$W$}-algebras.
\newblock {\em Mosc. Math. J.}, 12(3):633--666, 669--670, 2012.

\bibitem{Nakajima:1994}
Hiraku Nakajima et~al.
\newblock Instantons on {ALE} spaces, quiver varieties, and {K}ac-{M}oody
  algebras.
\newblock {\em Duke Mathematical Journal}, 76(2):365--416, 1994.

\bibitem{Nakajima:1998}
Hiraku Nakajima et~al.
\newblock Quiver varieties and {K}ac-{M}oody algebras.
\newblock {\em Duke Mathematical Journal}, 91(3):515--560, 1998.

\bibitem{Nakajima:03}
Hiraku Nakajima and K\=ota Yoshioka.
\newblock Lectures on instanton counting.
\newblock In {\em Algebraic structures and moduli spaces}, volume~38 of {\em
  CRM Proc. Lecture Notes}, pages 31--101. Amer. Math. Soc., Providence, RI,
  2004.

\bibitem{Nakajima:2005instanton}
Hiraku Nakajima and Kota Yoshioka.
\newblock Instanton counting on blowup. {I}. 4-dimensional pure gauge theory.
\newblock {\em Inventiones mathematicae}, 162(2):313--355, 2005.

\bibitem{Namikawa:2013poisson}
Yoshinori Namikawa.
\newblock Poisson deformations and birational geometry.
\newblock {\em arXiv preprint arXiv:1305.1698}, 2013.

\bibitem{Nekrasov:02}
Nikita~A. Nekrasov.
\newblock {c}.
\newblock {\em Adv. Theor. Math. Phys.}, 7(5):831--864, 2003.

\bibitem{Proudfoot:2014}
Nicholas Proudfoot and Travis Schedler.
\newblock Poisson--de {R}ham homology of hypertoric varieties and nilpotent
  cones.
\newblock {\em Selecta Math. (N.S.)}, 23(1):179--202, 2017.

\bibitem{Proudfoot:2004hyperkahler}
Nicholas~J Proudfoot.
\newblock Hyperkahler analogues of {K}ahler quotients.
\newblock {\em arXiv preprint math/0405233}, 2004.

\bibitem{Ruan:2000stringy}
Yongbin Ruan.
\newblock Stringy geometry and topology of orbifolds.
\newblock {\em arXiv preprint math/0011149}, 2000.

\bibitem{Ruan:2001cohomology}
Yongbin Ruan.
\newblock Cohomology ring of crepant resolutions of orbifolds.
\newblock {\em arXiv preprint math/0108195}, 2001.

\bibitem{Singletonthesis}
A.~Singleton.
\newblock The geometry and representation theory of superconformal quantum
  mechanics.
\newblock \url{https://www.repository.cam.ac.uk/handle/1810/260821}, 2016.
\newblock PhD thesis.

\bibitem{Singleton:2014}
Andrew Singleton.
\newblock {Superconformal quantum mechanics and the exterior algebra}.
\newblock {\em JHEP}, 06:131, 2014, {\emph{arXiv preprint hepth/1403.4933}}.

\bibitem{Thomason:1992}
Robert~W Thomason et~al.
\newblock Une formule de {L}efschetz en {K}-th{\'e}orie {\'e}quivariante
  alg{\'e}brique.
\newblock {\em Duke Mathematical Journal}, 68(3):447--462, 1992.

\end{thebibliography}

\end{document}